\documentstyle[12pt]{article}

\begin{document}

\baselineskip=24pt plus 1pt minus 1pt

\begin{center}

{\large \bf Odd--even staggering in octupole bands of actinides 
and rare earths: Systematics of ``beat'' patterns}

\bigskip\bigskip

{Dennis Bonatsos$^{1}$,
C.~Daskaloyannis$^2$,
S. B. Drenska$^3$,
N. Fotiades$^{4}$,
N. Minkov$^3$, 
P. P. Raychev$^3$,
R. P. Roussev$^{3*}$}
\bigskip

{$^{1}$ Institute of Nuclear Physics, N.C.S.R.
``Demokritos''}

{GR-15310 Aghia Paraskevi, Attiki, Greece}

{$^2$ Department of Physics, Aristotle University of
Thessaloniki}

{GR-54006 Thessaloniki, Greece }

{$^3$ Institute for Nuclear Research and Nuclear Energy, Bulgarian
Academy of Sciences }

{72 Tzarigrad Road, BG-1784 Sofia, Bulgaria}

{$^4$ Los Alamos National Laboratory, Los Alamos, New Mexico 87545}

\end{center}

\bigskip\bigskip
\centerline{\bf Abstract} 
\medskip

``Beat'' patterns are shown to appear in the octupole bands of several 
actinides and rare earths, their appearance being independent from 
the formula used in order to isolate and demonstrate them. It is shown 
that the recent formalism, making use of discrete approximations to 
derivatives of the transition energies (or of the energy levels) gives 
results consistent with the traditional formulae.  
In both regions it is seen that the first vanishing of the staggering
occurs at higher values of the angular momentum $I$ in nuclei exhibiting 
higher staggering at low $I$. Since these nuclei happen to be good 
rotators, the observed slow decrease of the amplitude of the staggering 
with increasing $I$ is in good agreement with the parameter independent 
predictions of the su(3) (rotational) limit of several algebraic models.  
In the actinides it has been found that within each 
series of isotopes the odd--even staggering exhibits minima at $N=134$
and $N=146$, while a local maximum is shown at $N=142$, these findings 
being in agreement with the recent suggestion of a secondary maximum 
of octupole deformation around $N=146$. 

\bigskip


Short title: Odd--even staggering in octupole bands 
\vfill
-----------------------------------

$^*$ Deceased

\newpage

\section{Introduction}

Negative parity bands with angular momenta 
$I^{\pi}=1^-$, 3$^-$, 5$^-$, 7$^-$, \dots, merging with the ground state 
bands (characterized by $I^{\pi}=0^+$, 2$^+$, 4$^+$, 6$^+$, \dots) 
into forming a single band with the level sequence $I^{\pi}=0^+$, 1$^-$, 
2$^+$, 3$^-$, 4$^+$, 5$^-$, \dots have been known to exist in the light 
actinides \cite{Schueler,Sood1}, and in particular in the Rn, Ra, and Th 
isotopes with neutron number $N=130$-138 \cite{Ahmad,Butler}, 
as well as in the rare earth region, and in 
particular in the Ba--Sm isotopes \cite{Phillips1,Sheline1,Phillips2,Sood2}
for a long time. These bands have been interpreted as corresponding to 
reflection asymmetric shapes, i.e. as indicating the existence of 
permanent octupole deformation \cite{Ahmad,Butler,Engel1,Engel2} of the 
corresponding nuclei, although alternative interpretations of these bands 
in terms of alpha clustering have been given \cite{Daley,Buck1,Buck2,Buck3}. 

An interesting quantity characterizing the structure of bands with level
sequence $I^{\pi}=0^+$, 1$^-$, 2$^+$, 3$^-$, \dots is the relative 
displacement of the odd levels with respect to the positions predicted 
for them by a fitting of the even levels. This quantity, usually 
called {\sl odd--even staggering} or $\Delta I=1$ {\sl staggering}, 
should vanish if the even levels and the odd levels form a single band. 

Traditionally it has been thought that the odd--even staggering in octupole 
deformed bands starts from a relatively high value at low $I$ and then 
gradually decreases down to zero, thus indicating the gradual formation 
of a band with a reflection asymmetric shape 
\cite{Schueler,Phillips1,Sheline1,NazOl}. However, using recent data 
in the actinide region \cite{Cocks1,Cocks2}, it has been found that 
in the light actinides the odd--even staggering exhibits a ``beating''
behavior \cite{Bon}. In other words, the quantity measuring the odd--even 
staggering does not stay near a vanishing value after reaching zero 
for the first time, but continues to raise and fall (in absolute value)
with increasing $I$, producing a figure resembling a sequence of beats. 

At this point the following questions are raised: 

1) Is there a ``beating'' behavior of the odd--even staggering in octupole 
bands of the rare earth region, similar to the one seen in the light 
actinides? 

2) If there is such a behavior, what are the systematics exhibited by these
``beats''? 

3) Is the ``beating'' behavior a real effect or is it an artifact of the 
mathematical method used? 

Furthermore, it has been recently argued \cite{Wieden,Sheline2} that 
in the actinides a second region of octupole 
deformation exists around the nuclei $^{238}$Pu 
and $^{240}$Pu, i.e. around $N=144$, 146. It is therefore of interest 
to study the systematics of odd--even staggering up to and around 
this region, since the following questions are raised:

1) Is there any systematic dependence of the odd--even staggering 
on the neutron number $N$ within each chain of isotopes? 

2) Is there any correlation between the amplitude of the odd--even 
staggering and the angular momentum value at which the vanishing 
of the staggering occurs? 

In section 2 of the present work the formalism will be briefly described 
and it will be subsequently applied to a few examples in the actinide region 
in section 3. In section 4 the consistency of several staggering formulae 
among themselves, as well as with the traditional formalism 
will be considered. The systematics of odd--even staggering in the actinide 
region will be examined in section 5, while section 6 will contain 
examples of ``beats'' in the rare earth region. The systematics of 
the odd--even staggering in the rare earth region will be considered 
in section 7, while section 8 will contain the conclusions drawn from 
the present results and plans for future work. 

\section{Formalism}

The odd--even staggering (or $\Delta I=1$ staggering) 
can be measured by the quantity \cite{Bon}
\begin{equation}
\Delta E_{1,\gamma}(I)= {1\over 16} (6 E_{1,\gamma}(I)-4 E_{1,\gamma}(I-1)
-4 E_{1,\gamma}(I+1) +E_{1,\gamma}(I-2) + E_{1,\gamma}(I+2) ),
\end{equation} %(1)
where the transition energies 
\begin{equation}
E_{1,\gamma}(I)= E(I+1)-E(I).
\end{equation} %(2)
are determined directly from experiment. This expression is analogous 
to the one used recently for the study of $\Delta I=2$ staggering 
\cite{Fli,Ced} in nuclear superdeformed bands.  
One of the main properties of eq. (1) is that it vanishes for 
\begin{equation}
E(I)= E_0+A I(I+1) + B [I(I+1)]^2,
\end{equation} %(3)
since it is a (normalized) discrete approximation of the fourth derivative 
of the function $E_{1,\gamma}(I)$ \cite{Abram}, i.e., essentially the fifth 
derivative of the function $E(I)$. Therefore eq. (1) is a sensitive 
probe of deviations from rotational behavior. 

\section{Examples of ``beats'' in the actinide region}

A couple of examples of staggering plots (for the nuclei $^{220}$Ra
and $^{224}$Ra) are shown in fig. 1 for illustrative purposes. 
More examples of staggering plots 
for several Rn, Ra, and Th isotopes can be found in ref. \cite{Bon}.
The following comments are in place: 

a) The term ``staggering'' is justified by the fact that the quantity 
$\Delta E_{1,\gamma}(I)$ exhibits values of alternating sign over 
extended regions of the angular momentum $I$. 

b) Odd--even staggering starts from relatively high values and is then 
decreasing with increasing angular momentum $I$. After reaching a 
vanishing value, staggering starts raising and then dropping again, 
giving an overall picture of ``beats'', which has been discussed 
in ref. \cite{Bon}. 

c) When the staggering reaches a vanishing value, a phase change occurs.  
In other words, while within the first ``beat'' positive values 
of staggering occur at even $I$ and negative values of staggering occur 
at odd $I$, indicating that in this region of $I$ values the odd levels 
are raised in relation to the corresponding even levels, within the 
second ``beat'' the opposite picture holds, i.e. positive values 
of staggering occur at odd $I$ and negative values of staggering 
occur at even $I$, indicating that in this region the odd levels 
are depressed in relation to the corresponding even levels. 
This phase change has been explained by the simple model of section V of ref. 
\cite{Bon}. It should be noticed that this phase change bears great 
similarity to the phase change occuring with signature inversion 
\cite{Riedinger,Lee} in odd--odd nuclei. 

d) Because of the phase change mentioned in c), when staggering reaches 
zero, two consecutive staggering points with the same sign (both positive 
or both negative) occur. In what follows we are going to use the angular
momenta of these two consecutive points in order to indicate the region
at which vanishing staggering occurs, labelling them by $I_{van}$.  

e) In fig. 1 the amplitude of the second ``beat'' appears to be smaller
than the amplitude of the first ``beat''. Similarly, the amplitude of the 
third ``beat'' appears to be smaller than the amplitude of the second 
``beat''. These observations are in rough agreement with the expectations 
\cite{Schueler,Phillips1}
in the early days of octupole deformed bands, that the staggering should 
decrease down to zero with increasing $I$ and then stay at a vanishing value,
since a single band with octupole deformation is then formed. The bands
shown in fig. 1, as well as several additional bands exhibited in ref. 
\cite{Bon}, indicate that staggering does not stay at a vanishing value 
after reaching zero for the first time, but the additional ``beats''
have smaller and smaller amplitudes, in qualitative agreement with the 
early expectations. 

\section{Alternative formulae} 

At this point one might wonder to which extent the characteristics of the 
odd--even staggering seen in the previous section, as well as in ref. 
\cite{Bon}, depend on the selection of the specific formula given in 
eq. (1). Indeed, many other choices are possible. 

Instead of the normalized discrete approximation of the fourth derivative 
of the function $E_{1,\gamma}(I)$, used in eq. (1), one can use the 
normalized discrete approximation of the second derivative of the 
same function, calculated using the values of the function at three points
\cite{Abram}
\begin{equation}
\Delta E _{23,\gamma}(I)= {1\over 4} (2 E_{1,\gamma}(I)-E_{1,\gamma}(I-1)
-E_{1,\gamma}(I+1)),
\end{equation}%(4)
or the normalized discrete approximation of the second derivative of 
the function $E_{1,\gamma}(I)$, calculated using the values of the function 
at five points \cite{Abram}
\begin{equation}
\Delta E_{25,\gamma}(I)= {1\over 64} (30 E_{1,\gamma}(I)-16 E_{1,\gamma}(I-1)
-16 E_{1,\gamma}(I+1) + E_{1,\gamma}(I-2) + E_{1,\gamma}(I+2)).
\end{equation}%(5)

The formulae given above use the transition energies defined in eq. (2). 
It is possible, however, to use in such formulae directly the energy 
levels $E(I)$. Before doing so, it is instructive to write eqs. (1), 
(4) and (5) in terms of the energy levels $E(I)$, by substituting eq. (2) 
into them. The results are 
\begin{equation}
\Delta E_{1\gamma}(I)= {1\over 16}[ E(I+3) -5 E(I+2) +10 E(I+1) -10 E(I)
+5 E(I-1) - E(I-2) ], 
\end{equation}%(6)
\begin{equation}
\Delta E_{23,\gamma}(I)= {1\over 4} [-E(I+2) + 3 E(I+1) -3 E(I) + E(I-1) ],
\end{equation}%(7)
\begin{equation}
\Delta E_{25,\gamma}(I) = {1\over 64} [ E(I+3) -17 E(I+2) +46 E(I+1) 
- 46 E(I) +17 E(I-1) - E(I-2) ]. 
\end{equation}%(8)

We proceed now to use in the derivative formulae the energy levels $E(I)$
in the place of the transition energies $E_{1,\gamma}(I)$. 
In this case the normalized discrete approximation of the 
fourth derivative of the function $E(I)$, calculated using the values 
of the function at five points \cite{Abram} will be 
\begin{equation}
\Delta E_{45,E}(I)={1\over 8}(6 E(I)-4 E(I-1) -4 E(I+1) +E(I-2)+E(I+2)),
\end{equation}%(9)
which is the analog of eq. (1) in the case in which we use directly the 
energy levels $E(I)$ and not the transition energies $E_{1,\gamma}(I)$. 
The difference in the normalization factor (by a factor of 2) is
justified by comparison to eq. (6), which contains the same quantity 
as eq. (1), but expressed in terms of the energy levels $E(I)$. 
The choice made here is that in both cases the numerator 
contains a number of terms which is twice the number appearing in the 
denominator. This choice guarantees that the accumulation of  errors 
of the quantities given in eqs. (6) and (9) will be handled on equal
footing. 

In a similar manner one can use the normalized discrete approximation of the 
second derivative of the function $E(I)$, calculated using the values of the 
function at three points \cite{Abram} 
\begin{equation}
\Delta E_{23,E}(I)= {1\over 2} (2 E(I)-E(I-1)-E(I+1)),
\end{equation}%(10)
which is similar to eq. (4), the normalization being fixed by comparison 
to eq. (7),  
or the normalized discrete approximation 
of the second derivative of the function $E(I)$, calculated using the values 
of the function at five points \cite{Abram}
\begin{equation}
\Delta E_{25,E}(I)={1\over 32}(30 E(I)- 16 E(I-1) -16 E(I+1)+E(I-2)+E(I+2)),
\end{equation}%(11)
which is an analog of eq. (5), the normalization being fixed by comparison 
to eq. (8).  

The notation  used above is of the type $\Delta E_{ij}$, where by $i$ 
we denote the order of the derivative, while by $j$ we denote the number 
of values of the function used for the calculation of
the derivative each time. 
According to this notation the quantity of eq. (1) could have been denoted 
as $\Delta E_{45,\gamma}(I)$, but we opted for keeping for it the 
previously \cite{Bon} used notation 
$\Delta E_{1,\gamma}(I)$.

It should be noticed at this point that eq. (10) is equivalent 
to the expression
\begin{equation}
\delta E(I)= E(I)-{1\over 2} [E(I-1)+E(I+1)],
\end{equation}%(12) 
which has been used in staggering studies in the past 
in octupole bands \cite{NazOl} and gamma bands \cite{PLB200}, as well
as in bands of odd--odd nuclei in connection with the effect of 
signature inversion \cite{Riedinger}.  

In addition it can be noticed that eq. (7) is equivalent (up to a factor 
of 2) to the expression 
$$\Delta E_S(I)= [E(I+1)-E(I)]-{E(I+2)-E(I+1)+E(I)-E(I-1)\over 2}$$
\begin{equation}
= {1\over 2} [-E(I+2)+3 E(I+1)-3 E(I-1)+E(I-2)],
\end{equation}%(13)
which has also been used in bands of odd--odd nuclei in relation 
with the effect of signature inversion \cite{Lee}.
The similarity of the phase change occuring in the present case,
when a ``beat'' is completed and the next one is starting, to the phase
change 
occuring at signature inversion in odd--odd nuclei has been already mentioned 
in comment c) of sec. 3. 

Since the early days of octupole deformation, the expression \cite{NazOl}
\begin{equation}
\Delta E_{NO}(I)=E(I)- {(I+1)E(I-1)+IE(I+1) \over 2I+1}
\end{equation}%(14)
has been widely used. This expression does not correspond to any derivative,
but it has the property to vanish for
\begin{equation}
E(I)=E_0+A I(I+1),
\end{equation}%(15)
giving in this way a measure of deviations from pure rotational behavior. 

From the theoretical point of view the main differences of the formulae 
given above can be described by considering the energy expression
\begin{equation}
E(I) = E_0 +  A' I + A I(I+1) + B [I(I+1)]^2.
\end{equation}%(16)
The following comments can be made: 

1) The expression $\Delta E_{1,\gamma}(I)$
``kills'' all terms of eq. (16), i.e. it vanishes if eq. (16) is 
substituted in it, while the expression $\Delta E_{45,E}(I)$
gives a constant coming from the last term of eq. (16). 
This is expected, since $\Delta E_{45,E}(I)$ corresponds 
to the fourth derivative of the energies $E(I)$, while $\Delta
E_{1,\gamma}(I)$
corresponds to the fourth derivative of the transition energies of eq. (2),
i.e. essentially to the fifth derivative of $E(I)$, as seen in eq. (6). 

2) The expressions $\Delta E_{23,E}(I)$ and $\Delta E_{25,E}(I)$ 
give a constant coming from the third term when 
the first three terms of eq. (16) are substituted in them, 
since they correspond to the second  
derivative of $E(I)$. The expressions $\Delta E_{23,\gamma}(I)$ and 
$\Delta E_{25,\gamma}(I)$ completely
``kill'' the first three terms of eq. (16),
since they correspond to the second derivative of the transition energies 
of eq. (2), i.e. essentially to the third derivative of $E(I)$, as seen in
eqs. (7) and (8). 

3) The expression $\Delta E_{NO}(I)$ ``kills'' the first and the third terms 
of eq. (16), but not the second one (which is of vibrational character),
as one can check by simple substitution. The main difference between 
$\Delta E_{NO}(I)$ and the other quantities is that $E_{NO}(I)$ does not 
correspond to any derivative.    
 
In order to demonstrate the similarities and differences between these 
formulae, we give in table 1 detailed results from their application 
in the set of data of $^{220}$Ra, a nucleus in which the third ``beat''
is reached. The following observations can be made: 

1) $\Delta E_{23,\gamma}(I)$ and $\Delta E_{25,\gamma}(I)$ give results 
which are almost identical, a fact expected, since both of these  quantities
represent the second derivative of the transition energies, calculated 
using the values of the function at a different number of points in each 
case. The results given by $\Delta E_{1,\gamma}(I)$ are also very similar
to the ones just mentioned, indicating that the fourth 
derivative gives almost the same results as the second derivative. 

2) $\Delta E_{23,E}(I)$ and $\Delta E_{25,E}(I)$ give almost identical 
results, which in addition are very similar to the results given by 
$\Delta E_{45,E}(I)$, indicating once more that the selection of the 
order of the derivative (second or fourth), as well as of the number 
of the values of the function used at each step of the calculation 
of the derivative (three or five) is not important.    

3) In table 1 the quantity $\Delta E_{1,\gamma}(I)$ is accompanied by the 
relevant errors, calculated from the errors of the experimental data 
\cite{Ra220}. It is clear that the errors are small in comparison 
to the magnitude of $\Delta E_{1,\gamma}(I)$. The errors for the 
rest of the quantities in table 1 are of similar magnitude and therefore 
are not reported. 

4) The overall sign difference between the quantities $\Delta
E_{1,\gamma}(I)$,
$\Delta E_{23,\gamma}(I)$, $\Delta E_{25,\gamma}(I)$ and the rest is due 
to the choice made in writing eq. (2), which is used in them. Using 
$E_{1,\gamma}(I)=E(I)-E(I+1)$ instead of eq. (2), one would have obtained 
for these quantities the same results with opposite signs. 

In order to check further the similarities and differences between the
various 
results, we have plotted in fig. 2a a representative of case 1) 
($\Delta E_{1,\gamma}(I)$), in which the transition energies are used, 
a representative of case 2) ($\Delta E_{45,E}(I)$), in which the energy 
levels are used, and $\Delta E_{NO}(I)$, which does not correspond 
to any derivative. The following comments can be made: 

1) All curves have exactly the same shape.
Errors are not shown, since they are smaller than 
the size of the symbols used for indicating the points, as seen in table 1.  
The sign difference has been explained in comment 4) above. 

2) In all cases the curves cross the zero axis at the same angular 
momentum values. This can also be seen in table 1, since the crossing 
of the zero corresponds, as explained above,  to a phase change, i.e.
to two consecutive values with the same sign.   

3) The traditional formula of eq. (14) \cite{NazOl} produces a staggering 
pattern very similar to the more recent formula of eq. (1) 
\cite{Bon,Fli,Ced}. In early work \cite{Schueler,Phillips1,NazOl}
the plots had been including the values of $\Delta E_{NO}(I)$ at 
odd $I$ only, thus producing a curve going down to zero with increasing $I$, 
instead of the first ``beat''. In some cases, in which more experimental 
data was available, the curve was continued 
with negative values for a few more odd values of $I$, corresponding 
to the beginning of the second ``beat'' in the current presentation. 
In other words, the first three ```beats'' seen in fig. 2a in the 
current presentation correspond in the case of the traditional description 
(using the formula of eq. (14) and plotting only the values at odd $I$) 
to a curve decreasing from positive values to negative values and then 
raising to positive values again. 

4) The previous point clarifies that staggering is not an artifact
constructed through the use of derivatives, but it is corresponding 
to a physical situation: Within the first and the third ``beat'' the negative 
parity band is lying higher than the position corresponding to 
an interpolation of the levels of the positive parity band, while  
within the second ``beat'' the negative parity band is lying lower 
than the position corresponding to the interpolation of the levels of the 
positive parity band. What the derivatives do, is to isolate and 
demonstrate this effect in a clear way. It is remarkable that eq. (14), 
which is not a derivative, performs this task equally well. 

In fig. 2b the same quantities are plotted in the case of $^{224}$Ra, 
leading to the same conclusions. Similar results have been obtained 
for the rest of the nuclei used in ref. \cite{Bon}, as well as for the 
nuclei used in the rest of the present work. 

From the above results the following conclusions can be drawn:

a) The staggering pattern does not depend on the selection of the 
derivative (second or fourth), or on the number of values of the 
function used each time for the calculation of the derivative (three or 
five). 

b) The staggering pattern does not depend on the use of the energy levels 
or of the transition energies in the relevant formulae. 

c) The staggering pattern is reconciled with the traditional description 
(using eq. (14) and plotting values only at odd $I$). It is related 
to the relative position of the levels of the negative parity band
with respect to the positions they should have had according to an 
interpolation of the levels of the positive parity band.  

d) The physical question remaining is why in the beginning
(within the first ``beat'', at low $I$) the negative parity band lies higher 
than where it should have been, later (within the second ``beat'') 
it lies lower than where it should have been, while at even higher 
angular momentum $I$ (within the third ``beat'') it lies again higher 
than where it should have been. 

\section{Systematics of odd--even staggering in the actinide region } 

In table 2 the first two values of the quantity of eq. (1), i.e.
$\Delta E_{1,\gamma} (3)$ and $\Delta E_{1,\gamma}(4)$, are reported 
for series of Rn, Ra, Th, U, and Pu isotopes, along with the pairs of angular 
momentum values 
at which the first vanishing of the odd--even staggering 
occurs, as explained in comment d) of section 3. The bands used in the table 
are free from backbendings/upbendings \cite{deVoigt}.
The following observations can be made: 

a) The Rn and Ra series of isotopes clearly show that minimum 
staggering occurs at $N=134$, the Th series of isotopes also being 
consistent with this observation. 

b) The U series of isotopes clearly 
demonstrates that maximum staggering occurs at $N=142$, the Th and Pu 
series of isotopes being consistent with this finding. 

c) The Pu 
series of isotopes clearly exhibit minimum staggering at $N=146$, 
the U series of isotopes being consistent with this finding. 

We conclude that low staggering (i.e. strong octupole correlations) 
is observed around $N=134$ (the well known area of octupole deformation 
in the light actinides \cite{Ahmad,Butler}) 
and around $N=146$ (the recently suggested 
region of octupole deformation in heavier actinides \cite{Wieden,Sheline2}), 
while relatively 
high staggering (weak octupole correlations) is observed around $N=142$.
These findings are consistent with the suggestions of ref. \cite{Sheline2},
which have been based on systematics of the energies of the lowest $1^-$ 
states, as well as on systematics of the hindrance factors for 
alpha decay. 

The data in table 2 suggest in addition that the lower the initial staggering
(i.e. the staggering at $I=3$, 4), 
the lower the angular momentum value at which the staggering vanishes. 
However, this conclusion is based only on some of the Rn, Ra, and Th
isotopes, for which the vanishing of the staggering has been reached 
experimentally, while in the Pu and U isotopes, as well as in several 
of the heavier Th isotopes, the vanishing of the staggering has not 
been reached experimentally yet.  

The suggestion that the lower the initial staggering, the lower the angular 
momentum value at which the staggering vanishes, is also supported by 
fig. 3, in which the odd--even staggering for the Th and U series 
of isotopes, calculated from eq. (1),  is shown. 
It is clear that the curves enveloping the staggering
pattern of each nucleus (not drawn in the figures for reasons of clarity)
do not cross each other, suggesting in this way that the nucleus 
possessing the lowest initial staggering will reach vanishing staggering 
first (i.e. at lower angular momentum), while the nucleus possessing 
the highest initial staggering will reach vanishing staggering last 
(i.e. at higher angular momentum), provided that no backbending
\cite{deVoigt} will appear meanwhile. A similar conclusion can be drawn 
for the Pu series of isotopes from fig. 2 of Ref. \cite{Wieden}, where 
eq. (14) has been used in the traditional way.  

In table 3 the $\Delta E_{1,\gamma}(I)$ values for five of the nuclei 
shown in fig. 3 are given as examples, 
together with their uncertainties, calculated 
from the experimental errors accompanying the data on the energy levels.
It is clear that in all cases the uncertainties are small, i.e. smaller than
the size of the symbols used in fig. 3. 

The following comments are here in place: 

a) In table 2 it is clear that the nuclei for which the first vanishing 
of the staggering has been reached experimentally, and therefore the relevant 
$I_{van}$ values are reported in the table, are vibrational 
(with $R_4=E(4)/E(2)$ ratios in the range
$2\leq R_4 \leq 2.4$), transitional (characterized by 
$2.4\leq R_4 \leq 3$), or rotational ($3\leq R_4\leq 10/3$) but with 
$R_4$ up to roughly 3.14~. In contrast, the Ra, Th, U, and Pu isotopes 
for which long bands are known, but the first vanishing of the staggering 
has not been reached yet, are all rotational, with $R_4> 3.2$.  
It seems therefore that the decrease of the staggering with increasing 
$I$ is in general slower in good rotators.

b) As a result of a), the good rotators shown in fig. 3 give the impression 
that their staggering is in rough agreement with the predictions 
of the su(3) (rotational) limit of several algebraic models, as the 
spf-Interacting Boson Model \cite{Engel1}, 
the spdf-Interacting Boson Model \cite{Engel1,Engel2}, 
the Vector Boson Model \cite{VBM1,VBM2,VBM3}, 
the Nuclear Vibron Model \cite{Daley}, 
the details of which 
have been described in ref. \cite{Bon} and need not be repeated here. 
All these models 
in their su(3) limits predict staggering of constant amplitude, 
the predictions being parameter independent, thus providing quite strict 
tests for the validity of these models.  
Fig. 3 shows that the above mentioned algebraic models pass this parameter 
independent test quite successfully. 

As far as the ``beat'' patterns are concerned, it should be mentioned 
that quite satisfactory results can be obtained within the recently 
suggested \cite{Rad97,Rad98} extension to negative parity states
of the Coherent State Model \cite{Rad81,Rad82}. Theoretical predictions 
for the octupole bands of $^{218-222}$Rn and $^{218-226}$Ra have 
been given in refs \cite{Rad97,Rad98}, by fitting the relevant Hamiltonian,
which possesses 7 free parameters, to the experimental data. Plugging 
the theoretical predictions of refs \cite{Rad97,Rad98} into eq. (1)
one can easily see that in all of the just mentioned Rn and Ra isotopes 
the first vanishing of the staggering is reproduced correctly. Furthermore 
the second vanishing of the staggering is reproduced correctly in 
$^{224}$Ra, while in $^{220}$Ra and $^{226}$Ra it is not reproduced
by the specific fits given in refs \cite{Rad97,Rad98}. However it is 
interesting to examine if this can be achieved with different parameter sets. 

\section{Examples of ``beats'' in the rare earth region} 

After discussing in some detail the actinide region, we now turn our
attention 
to the rare earth region. 
Staggering plots for 4 nuclei for which staggering reaches a vanishing 
value and goes on beyond it are shown in fig. 4. The following comments apply:

a) The staggering patterns seen here resemble the ones seen in sec. 3, 
but they are shorter. 
In $^{150}$Sm (fig. 4(c)) the second ``beat'' is almost complete, 
in $^{144}$Ba (fig. 4(a)) about half of the second ``beat'' is seen, 
while in $^{146}$Ba (fig. 4(b)) and $^{154}$Dy (fig. 4(d))
only the beginning of the second ``beat'' is observed. 

b) The best ``beat'' patterns 
are seen in the $N=88$ isotones ($^{144}$Ba, $^{150}$Sm, $^{154}$Dy),
which are known to be among the best examples of nuclei showing octupole 
deformation in the rare earth region \cite{Sheline1}. 

In table 4 the $\Delta E_{1,\gamma}(I)$ values for $^{150}$Sm are given 
as an example, 
together with their uncertainties, as calculated from the 
experimental errors accompanying the data on the energy levels. 
It is clear that the uncertainties are smaller than the size of the symbols 
used in fig. 4 

\section{Systematics of odd--even staggering in the rare \hfill\break 
earth region}

We proceed to a study of the systematics of odd--even staggering in the 
rare earth region, similar to the one given in sec. 5 for the actinides. 

In fig. 5 several Sm, Gd, Dy, and Er isotopes, for which the first 
vanishing 
of the staggering is not reached within the angular momentum region 
observed, are shown. In all cases it is clear that the curves enveloping 
the staggering patterns of the various isotopes (not shown in the figure 
for reasons of clarity) do not cross each other. In other words, the 
higher the initial staggering (at low $I$), the higher the angular 
momentum value at which the first vanishing of the staggering will occur
(provided that no backbending \cite{deVoigt} will occur meanwhile).   
This is in agreement with what we have seen in sec. 5 for the actinides. 

In table 4, which is an analog of table 3, the $\Delta E_{1,\gamma}(I)$ 
values for five of the nuclei shown in fig. 5 are given as examples, 
together with 
their uncertainties, as calculated from the experimental errors 
accompanying the data on the energy levels. It is clear that in all cases 
the uncertainties are smaller than the size of the symbols used in fig. 5.  
 
In table 5, which is an analog of table 2, 
the first two values of the quantity of eq. (1), i.e.
$\Delta E_{1,\gamma} (3)$ and $\Delta E_{1,\gamma}(4)$, are reported 
for several Ba, Ce, Nd, Sm, Gd, Dy, and Er isotopes (including all the 
isotopes appearing in figs 4 and 5), 
along with the pairs of angular momentum values 
at which the first vanishing of the odd--even staggering 
occurs, as explained in comment d) of sec. 3. 
The bands used in table 5
(as well as in figs 4 and 5) 
are free from backbendings \cite{deVoigt}, at least within the angular 
momentum regions used. 
The following observations can be made: 

a) The Nd, Sm, and Gd series of isotopes show that minimum 
staggering occurs at $N=86$, the Dy series of isotopes also showing
in the same direction.  

b) The Ba, Nd, and Sm
series of isotopes show that the lower the initial staggering
(i.e. the staggering at $I=3$, 4), 
the lower the angular momentum value at which the staggering vanishes. 
This observation is consistent with the suggestions of fig. 5, 
which are based on Sm, Gd, Dy, and Er isotopes.  

c) The nuclei for which the first vanishing of the staggering is reached 
(and therefore the relevant $I_{van}$ values are reported in table 5)
are either vibrational (characterized by $R_4=E(4)/E(2)$ ratios 
in the range $2\leq R(4) \leq 2.4$) or transitional (having 
$2.4\leq R(4) \leq 3$), while most of the nuclei shown in fig. 5, 
for which the first vanishing of the staggering is not reached up 
to quite high $I$, are purely rotational, having $3\leq R(4) \leq 10/3$
(with the notable exception of $^{152}$Gd). 

d) Furthermore, the rotational nuclei included  
in fig. 5 show a slow decrease of the amplitude of the staggering
with increasing $I$, being therefore quite in agreement with   
several algebraic models \cite{Engel1,Engel2,Daley,VBM1,VBM2,VBM3},
which in their su(3) (rotational) limits
predict staggering of constant amplitude. The predictions 
of these models (spf-Interacting Boson Model \cite{Engel1}, 
spdf-Interacting Boson Model \cite{Engel1,Engel2}, Vector Boson Model
\cite{VBM1,VBM2,VBM3}, Nuclear Vibron Model \cite{Daley}) 
have been described in detail in ref. \cite{Bon}
and need not be repeated here. It is worth mentioning, however, that 
these predictions of the just mentioned models are parameter independent,
thus providing a quite strict test of the validity of the models,
which they pass quite successfully. 

e) The observations discussed in comments b), c), and d) are similar 
to the ones made in the actinide region (see sec. 5). 

It will be interesting to examine if the recently proposed \cite{Rad97,Rad98}
extension to negative parity states of the Coherent State Model 
\cite{Rad81,Rad82}, which has been quite successful in describing the 
``beat'' patterns in the actinide region, as already mentioned in sec. 5,  
can also reproduce the ``beat'' patterns in the rare earth region. 

\section{Conclusion}

In this paper the odd--even staggering in the octupole bands of 
several even-even actinides and rare earths 
and the appearance of ``beat'' patterns 
have been considered. The main conclusions are listed here. 

a) ``Beat'' patterns, similar to the ones reported in ref. \cite{Bon}
for the light actinides,
are seen in both the actinide and the rare earth regions. 

b) The odd--even staggering and the ``beat'' patterns do not depend 
on the choice of the formula used for isolating and demonstrating 
the effect. Several different formulae, corresponding to discrete 
approximations of derivatives of the transition energies, or of the 
energy levels themselves, give results which are perfectly consistent
among themselves. 

c) The recently used formalism, which makes use of discrete approximations 
of derivatives, gives results perfectly consistent with the ones provided 
by the traditional formula of Nazarewicz and Olanders \cite{NazOl}.  

d) In both the actinide and the rare earth regions it is seen that 
the higher the initial staggering (i.e. the staggering at low angular 
momentum $I$), the higher the angular momentum value at which the 
staggering will reach a vanishing value. 

e) In both regions it is also seen that in vibrational and transitional 
nuclei the first vanishing of the staggering occurs soon, while 
in rotational nuclei the first vanishing of the staggering is not reached 
up to much higher angular momentum $I$, giving the impression of 
staggering with slowly decreasing amplitude, in rough agreement with 
several algebraic models predicting staggering of constant amplitude 
in their su(3) (rotational) limits.  

f) In both regions it is also seen that the amplitude of the second ``beat''
appears to be smaller than the amplitude of the first ``beat''. In the 
actinides it is further observed that subsequent ``beats'' have even 
smaller amplitudes. These observations are in rough agreement with 
the early expectations that after the formation of a single band with 
octupole deformation is reached as $I$ is increasing, the staggering 
from there on should remain at or near a vanishing value. 

g) In the actinide region, in which several 
Rn, Ra, Th, U, and Pu isotopes have been considered, it has been found that 
within each series of isotopes the odd--even staggering exhibits minima 
around $N=134$ and $N=146$, while a local maximum appears at $N=142$, 
in agreement with the recent suggestion \cite{Wieden,Sheline2}
of a secondary maximum of octupole deformation around $N=142$. 
 
h) In the rare earth region, minimum staggering occurs in the $N=86$, 88 
isotones, which show the best octupole bands and also the best ``beat''
patterns in this region. 

The simple model of ref. \cite{Bon} explains the main features of staggering, 
except the decrease of the maximum amplitude of subsequent ``beats'' 
with increasing $I$, which remains an open problem, along with 
a microscopic interpretation of the ``beat'' patterns in both the rare 
earth and the actinide regions. 

{\bf Acknowledgements} 

One of the authors (NF) acknowledges support by the U.S. Department of 
Energy under Contract Nos. W-7405-ENG-36 (LANL). 
Another author (PPR) acknowledges support by the Bulgarian Ministry 
of Science and Education under Contract No. $\Phi$-547.   
Another author (NM) has been supported by the Bulgarian National Fund 
for Scientific Research under Contract No. MU-F-02/98.

\newpage 

%%%%%%%%%%%%%%%%%%%%%%%%%%%%%%%%%%%%%%%%%%%%%%%%%%%%%%%%%%%%%%%%%%%%%%%%
%%%%%%%%%%%%%%%%%%%%%%% Table 1 %%%%%%%%%%%%%%%%%%%%%%%%%%%%%%%%%%%%%%%%
\begin{table}

\caption{Odd--even staggering quantities 
$\Delta E_{1,\gamma}(I)$ (eq. (1),  column 2), 
$\Delta E_{23,\gamma}(I)$ (eq. (4), column 4), 
$\Delta E_{25,\gamma}(I)$ (eq. (5), column 5), 
$\Delta E_{45,E}(I)$ (eq. (9), column 6), 
$\Delta E_{23,E}(I)$ (eq. (10), column 7), 
$\Delta E_{25,E}(I)$ (eq. (11), column 8), and 
$\Delta E_{NO}(I)$ (eq. (14), column 9), 
all in keV, listed with increasing angular momentum $I$ (column 1)
for the nucleus $^{220}$Ra. Experimental data have been taken from ref. 
\cite{Ra220}. In column 3 the uncertainties of $\Delta E_{1,\gamma}(I)$, 
occuring from the uncertainties of the experimental data given 
in ref. \cite{Ra220} are listed, in keV.}

\bigskip

\centering
\begin{tabular}{r r r r r r r r r }
\hline
$I$ & $\Delta E_{1,\gamma}$ & error & $\Delta E_{23,\gamma}$ &
$\Delta E_{25,\gamma}$ & $-\Delta E_{45,E}$ & $-\Delta E_{23,E}$ &
$-\Delta E_{25,E}$ & $-\Delta E_{NO}$  \\
\hline 
 2 &        &     &  222.5 &        &       & 265.1 &       &  259.0 \\
 3 & -165.5 & 0.5 & -162.2 & -163.0 &-192.3 &-179.9 &-183.0 & -196.4 \\
 4 &  117.0 & 0.6 &  115.1 &  115.6 & 138.6 & 144.4 & 143.0 &  135.5 \\
 5 &  -78.4 & 0.7 &  -75.8 &  -76.4 & -95.4 & -85.7 & -88.1 &  -98.4 \\
 6 &   48.2 & 0.7 &   47.1 &   47.3 &  61.4 &  65.8 &  64.7 &   56.6 \\
 7 &  -24.3 & 0.8 &  -22.8 &  -23.2 & -34.9 & -28.4 & -30.0 &  -38.8 \\
 8 &    5.3 & 0.8 &    4.5 &    4.7 &  13.7 &  17.2 &  16.3 &    8.6 \\
 9 &    9.5 & 0.9 &   10.5 &   10.2 &   3.0 &   8.2 &   6.9 &   -0.8 \\
10 &  -21.3 & 1.0 &  -21.8 &  -21.7 & -16.1 & -12.8 & -13.6 &  -20.7 \\
11 &   30.5 & 1.0 &   31.1 &   30.9 &  26.4 &  30.9 &  29.7 &   22.8 \\
12 &  -37.7 & 1.0 &  -38.0 &  -37.9 & -34.6 & -31.3 & -32.1 &  -38.7 \\
13 &   43.2 & 1.0 &   43.7 &   43.6 &  40.9 &  44.8 &  43.8 &   37.4 \\
14 &  -47.0 & 1.0 &  -47.4 &  -47.3 & -45.5 & -42.6 & -43.3 &  -49.5 \\
15 &   49.1 & 1.1 &   49.7 &   49.5 &  48.5 &  52.1 &  51.2 &   45.4 \\
16 &  -49.5 & 1.2 &  -49.9 &  -49.8 & -49.8 & -47.2 & -47.8 &  -53.7 \\
17 &   48.0 & 1.2 &   48.6 &   48.5 &  49.3 &  52.6 &  51.8 &   46.3 \\
18 &  -44.4 & 1.2 &  -44.9 &  -44.8 & -46.8 & -44.6 & -45.1 &  -50.7 \\
19 &   38.7 & 1.2 &   39.3 &   39.1 &  42.1 &  45.3 &  44.5 &   39.4 \\
20 &  -31.0 & 1.2 &  -31.2 &  -31.2 & -35.3 & -33.4 & -33.8 &  -39.2 \\
21 &   21.9 & 1.3 &   22.2 &   22.1 &  26.7 &  29.1 &  28.5 &   23.6 \\
22 &  -12.1 & 1.4 &  -12.1 &  -12.1 & -17.1 & -15.3 & -15.7 &  -20.8 \\
23 &    1.9 & 1.4 &    2.1 &    2.0 &   7.1 &   9.0 &   8.5 &    3.8 \\
24 &    8.6 & 1.4 &    8.5 &    8.5 &   3.2 &   4.8 &   4.4 &   -0.4 \\
25 &  -19.5 & 1.4 &  -19.4 &  -19.4 & -13.9 & -12.2 & -12.6 &  -17.0 \\
26 &   30.8 & 1.5 &   30.8 &   30.8 &  25.1 &  26.6 &  26.2 &   21.6 \\ 
27 &  -41.7 & 1.6 &  -42.1 &  -42.0 & -36.4 & -35.1 & -35.4 &  -39.7 \\
28 &   51.7 & 1.7 &   51.9 &   51.9 &  47.0 &  49.1 &  48.5 &   44.3 \\
29 &        &     &  -61.0 &        & -56.5 & -54.8 & -55.2 &  -59.2 \\
30 &        &     &        &        &       &  67.3 &       &        \\
\hline
\end{tabular}
\end{table}

\newpage 

%%%%%%%%%%%%%%%%%%%%%%%%%%%%%%%%%%%%%%%%%%%%%%%%%%%%%%%%%%%%%%%%%%%%%%
%%%%%%%%%%%%%%%%%%% Table 2  %%%%%%%%%%%%%%%%%%%%%%%%%%%%%%%%%%%%%%%%

\begin{table}

\caption{Odd--even staggering quantities $\Delta E_{1,\gamma}(3)$
(column 3) and $\Delta E_{1,\gamma}(4)$ (column 4),
in keV, calculated from eq. (1), for octupole 
bands of several actinides (listed in column 1).
The $R_4=E(4)/E(2)$ ratios of these nuclei are given in column 6.  
The couple of angular momentum values at which
the first vanishing of the staggering occurs (see text for further discussion)
is indicated by $I_{van}$ (column 5), in the cases in which it is known 
experimentally. Data have been taken from the references indicated in 
column 2.}
\bigskip

\centering
\begin{tabular}{ c c r r c c}
\hline
nucleus  & Ref. & $\Delta E_{1,\gamma}(3)$ & $\Delta E_{1,\gamma}(4)$ & 
$I_{van}$ & $R_4$ \\
\hline
                        &       &        &       &        &       \\
$^{218}_{86}$Rn$_{132}$ &[16,17]& -322.2 & 241.1 &        & 2.014 \\
$^{220}_{86}$Rn$_{134}$ &[16,17]& -257.6 & 188.5 & 10, 11 & 2.214 \\ 
$^{222}_{86}$Rn$_{136}$ &[16,17]& -296.9 & 230.3 & 10, 11 & 2.408 \\
                        &       &        &       &        &       \\
$^{220}_{88}$Ra$_{132}$ & [24]  & -165.5 & 117.0 &  8, 9  & 2.298 \\
$^{222}_{88}$Ra$_{134}$ &[16,17]& -103.8 &  71.8 &  7, 8  & 2.715 \\
$^{224}_{88}$Ra$_{136}$ &[16,17]& -117.1 &  89.5 &  8, 9  & 2.970 \\
$^{226}_{88}$Ra$_{138}$ &[16,17]& -177.0 & 152.3 & 12, 13 & 3.127 \\
$^{228}_{88}$Ra$_{140}$ & [17]  & -397.1 & 369.4 &        & 3.207 \\
                        &       &        &       &        &       \\
$^{224}_{90}$Th$_{134}$ & [28]  & -110.0 &  78.7 &  8, 9  & 2.896 \\
$^{226}_{90}$Th$_{136}$ & [29]  & -155.3 & 133.0 & 16, 17 & 3.136 \\
$^{228}_{90}$Th$_{138}$ & [30]  & -271.8 & 253.2 &        & 3.235 \\
$^{230}_{90}$Th$_{140}$ & [17]  & -456.3 & 437.9 &        & 3.271 \\
$^{232}_{90}$Th$_{142}$ & [17]  & -667.2 & 650.9 &        & 3.283 \\
$^{234}_{90}$Th$_{144}$ & [17]  &        &       &        & 3.308 \\
                        &       &        &       &        &       \\
$^{230}_{92}$U$_{138}$  & [31]  & -325.6 & 313.2 &        & 3.277 \\
$^{232}_{92}$U$_{140}$  & [32]  & -528.8 & 518.9 &        & 3.291 \\
$^{234}_{92}$U$_{142}$  & [33]  & -759.1 & 752.5 &        & 3.296 \\
$^{236}_{92}$U$_{144}$  & [34]  & -646.2 & 632.2 &        & 3.304 \\
$^{238}_{92}$U$_{146}$  & [35]  & -632.3 & 614.2 &        & 3.305 \\
                        &       &        &       &        &       \\
$^{238}_{94}$Pu$_{144}$ & [35]  & -565.8 &       &        & 3.311 \\
$^{240}_{94}$Pu$_{146}$ &[36,37]& -554.5 &       &        & 3.311 \\
$^{242}_{94}$Pu$_{148}$ & [38]  & -733.5 &       &        & 3.307 \\
                        &       &        &       &        &       \\
\hline                                    
\end{tabular}
\end{table}

\newpage

%%%%%%%%%%%%%%%%%%%%%%%%%%%%%%%%%%%%%%%%%%%%%%%%%%%%%%%%%%%%%%%%%%%%%%%%
%%%%%%%%%%%%%%%%%%%%%%%%%%%%% Table 3 %%%%%%%%%%%%%%%%%%%%%%%%%%%%%%%%%%%

\begin{table} 

\caption{Odd--even staggering quantity $\Delta E_{1,\gamma}(I)$ (eq. (1), 
in keV) listed with increasing angular momentum $I$ for the actinides 
$^{226}$Th \cite{Th226}, $^{228}$Th \cite{Th228}, $^{232}$U \cite{U232}, 
$^{236}$U \cite{U236}, and $^{238}$U \cite{U238}. The uncertainties, 
calculated from the experimental errors given in the 
same references, are given in parentheses, also in keV.}

\bigskip

\centering
\begin{tabular}{r r r r r r}
\hline
$I$ &$^{226}$Th & $^{228}$Th & $^{232}$U  & $^{236}$U  & $^{238}$U  \\
\hline
   &            &            &            &            &            \\   
 3 &-155.3 (0.2)&-271.8 (0.0)&-528.8 (0.2)&-646.2 (0.3)&-632.3 (0.3)\\
 4 & 133.0 (0.3)& 253.2 (0.1)& 518.9 (0.4)& 632.2 (0.6)& 614.2 (0.5)\\
 5 &-110.9 (0.4)&-233.0 (0.2)&-507.7 (0.6)&-616.2 (0.8)&-593.6 (0.6)\\
 6 &  90.4 (0.4)& 212.0 (0.4)& 495.6 (0.8)& 598.7 (0.9)& 570.9 (0.7)\\
 7 & -71.8 (0.4)&-190.7 (0.5)&-482.8 (0.8)&-580.4 (1.0)&-546.7 (0.8)\\
 8 &  55.7 (0.5)& 169.6 (0.6)& 469.7 (0.9)& 561.6 (1.2)& 521.5 (0.9)\\
 9 & -42.1 (0.6)&-148.9 (0.7)&-456.5 (1.0)&-542.8 (1.4)&-496.0 (1.0)\\
10 &  30.9 (0.7)& 128.9 (0.8)&            & 524.4 (1.6)& 470.4 (1.0)\\
11 & -22.0 (0.8)&-109.5 (0.8)&            &-506.7 (1.9)&-445.0 (1.1)\\
12 &  15.1 (0.9)&  90.9 (0.9)&            & 489.9 (2.1)& 420.2 (1.3)\\
13 &  -9.9 (1.0)& -73.3 (1.0)&            &-473.9 (2.4)&-396.0 (1.4)\\
14 &   6.1 (1.0)&  56.5 (1.0)&            & 458.9 (2.6)& 372.9 (1.5)\\
15 &  -3.3 (1.1)& -40.1 (1.1)&            &-445.5 (2.9)&-351.0 (1.6)\\
16 &   1.3 (1.2)&            &   & 433.7$\enskip\qquad$& 330.5 (1.9)\\
17 &   0.3 (1.3)&            &   &-423.5$\enskip\qquad$&-311.7 (2.3)\\
18 &            &            &            &            & 294.8 (2.6)\\
19 &            &            &            &            &-280.7 (2.9)\\
20 &            &            &            &            & 269.5 (3.1)\\
21 &            &            &            &            &-260.5 (3.4)\\
   &            &            &            &            &            \\
\hline
\end{tabular}
\end{table}

\newpage

%%%%%%%%%%%%%%%%%%%%%%%%%%%%%%%%%%%%%%%%%%%%%%%%%%%%%%%%%%%%%%%%%%%%%%%%
%%%%%%%%%%%%%%%%%%%%%%%% Table 4 %%%%%%%%%%%%%%%%%%%%%%%%%%%%%%%%%%%%%%

\begin{table}

\caption{Same as table 3, but for the rare earths $^{150}$Sm \cite{A150}, 
$^{152}$Sm \cite{A152}, $^{152}$Gd \cite{A152}, 
$^{154}$Gd \cite{A154}, $^{156}$Gd \cite{A156}, and $^{162}$Er \cite{A162}.}

\bigskip

\centering
\begin{tabular}{r r r r r r r}
\hline
$I$&$^{150}$Sm &$^{152}$Sm  &$^{152}$Gd  &$^{154}$Gd  &$^{156}$Gd   &
$^{162}$Er\\
\hline
  &            &            &            &            &             &\\
 3&            &-781.3 (0.0)&-587.3 (0.1)&-983.4 (0.0)&-1070.8 (0.0)&
-1112.0 (0.1)\\
 4& 389.8 $\qquad$     & 725.3 (0.1)& 510.4 (0.2)& 909.5 (0.0)& 1012.8 (0.0)&
 1026.6 (0.2)\\
 5&-311.2 $\qquad$     &-672.3 (0.1)&-464.5 (0.3)&-842.9 (0.1)& -955.9 (0.1)&
 -940.5 (0.2)\\
 6& 245.1 (0.1)& 623.8 (0.1)& 420.2 (0.4)& 783.3 (0.2)&  901.1 (0.1)&
  854.7 (0.3)\\
 7&-185.6 (0.2)&-579.8 (0.1)&-376.6 (0.6)&-729.9 (0.3)& -849.4 (0.1)&
 -769.9 (0.3)\\
 8& 129.8 (0.4)& 540.0 (0.2)& 333.3 (0.9)& 681.9 (0.4)&  801.2 (0.1)&
  687.0 (0.3)\\
 9& -77.8 (0.7)&-504.0 (0.2)&-290.4 (1.3)&-638.3 (0.4)& -757.1 (0.1)&
 -606.6 (0.3)\\
10&  30.5 (1.1)& 471.1 (0.2)& 248.2 (1.7)& 598.0 (0.4)&  717.4 (0.1)&
  529.9 (0.3)\\
11&  11.3 (1.5)&-440.5 (0.3)&-208.1 (2.0)&-560.0 (0.4)& -681.9 (0.1)&
 -459.1 (0.3)\\
12& -45.3 (1.8)&            & 172.5 (2.3)& 523.3 (0.3)&             &\\
13&  65.9 (2.0)&            &-146.0 (2.7)&-486.8 (0.3)&             &\\
14& -70.0 (2.4)&            & 131.1 (3.0)& 451.2 (0.4)&             &\\
15&  56.4 (2.7)&            &            &-416.5 (0.4)&             &\\
16& -25.3 (3.0)&            &            & 380.9 (0.5)&             &\\
17& -18.1 (3.2)&            &            &-346.3 (0.5)&             &\\
18&  64.9 (3.4)&            &            & 318.4 (0.6)&             &\\
19&            &            &            &-297.4 (0.7)&             &\\
20&            &            &            & 279.0 (0.8)&             &\\
21&            &            &            &-260.9 (0.9)&             &\\
22&            &            &            & 242.3 (1.0)&             &\\
23&            &            &            &-225.0 (1.1)&             &\\
  &            &            &            &            &             &\\
\hline
\end{tabular}
\end{table}

\newpage

%%%%%%%%%%%%%%%%%%%%%%%%%%%%%%%%%%%%%%%%%%%%%%%%%%%%%%%%%%%%%%%%%%%%%%
%%%%%%%%%%%%%%%%%%% Table 5  %%%%%%%%%%%%%%%%%%%%%%%%%%%%%%%%%%%%%%%%

\begin{table}

\caption{Same as table 2, but for octupole bands of several rare earths.}

\bigskip

\centering
\begin{tabular}{ c c r r c c}
\hline
nucleus     & Ref. & $\Delta E_{1,\gamma}(3)$ & $\Delta E_{1,\gamma}(4)$ & 
$I_{van}$ & $R_4$ \\
\hline
                       &        &        &       &        &       \\
$^{144}_{56}$Ba$_{88}$ & [47]   & -442.6 & 351.8 &  9, 10 & 2.662 \\
$^{146}_{56}$Ba$_{90}$ & [47]   & -442.2 & 349.5 &  9, 10 & 2.835 \\
$^{148}_{56}$Ba$_{92}$ & [48]   & -469.1 & 396.9 &        & 2.984 \\
                       &        &        &       &        &       \\
$^{146}_{58}$Ce$_{88}$ & [49]   & -456.2 & 336.6 &  8, 9  & 2.586 \\ 
                       &        &        &       &        &       \\
$^{146}_{60}$Nd$_{86}$ & [49]   & -399.3 & 211.0 &  5, 6  & 2.297 \\
$^{148}_{60}$Nd$_{88}$ & [50]   & -432.2 & 302.5 &  8, 9  & 2.493 \\
$^{150}_{60}$Nd$_{90}$ & [51]   & -666.3 & 615.6 &        & 2.929 \\
$^{152}_{60}$Nd$_{92}$ & [52]   &-1084.7 &       &        & 3.263 \\
                       &        &        &       &        &       \\
$^{148}_{62}$Sm$_{86}$ & [48]   & -287.0 & 124.8 &  5, 6  & 2.145 \\
$^{150}_{62}$Sm$_{88}$ & [51]   &        & 389.8 & 10, 11 & 2.316 \\
$^{152}_{62}$Sm$_{90}$ & [52]   & -781.3 & 725.3 &        & 3.009 \\
$^{154}_{62}$Sm$_{92}$ & [53]   & -832.9 & 801.9 &        & 3.254 \\
                       &        &        &       &        &       \\
$^{150}_{64}$Gd$_{86}$ & [51]   & -203.0 &  48.8 &        & 2.019 \\
$^{152}_{64}$Gd$_{88}$ & [52]   & -587.3 & 510.4 &        & 2.194 \\
$^{154}_{64}$Gd$_{90}$ & [53]   & -983.4 & 909.5 &        & 3.015 \\ 
$^{156}_{64}$Gd$_{92}$ & [54]   &-1070.8 &1012.8 &        & 3.239 \\ 
$^{158}_{64}$Gd$_{94}$ & [55]   &-1246.1 &       &        & 3.288 \\
$^{160}_{64}$Gd$_{96}$ & [56]   &-1118.0 &1077.4 &        & 3.302 \\
                       &        &        &       &        &       \\
$^{154}_{66}$Dy$_{88}$ & [53]   & -679.0 & 596.2 & 22, 23 & 2.233 \\
$^{156}_{66}$Dy$_{90}$ & [57]   &-1069.4 & 992.7 &        & 2.932 \\
$^{158}_{66}$Dy$_{92}$ & [55]   &-1171.9 &       &        & 3.206 \\
$^{162}_{66}$Dy$_{96}$ & [58]   &-1177.4 &1140.6 &        & 3.294 \\
                       &        &        &       &        &       \\
$^{162}_{68}$Er$_{94}$ & [58]   &-1112.0 &1026.6 &        & 3.230 \\
$^{164}_{68}$Er$_{96}$ & [59]   &-1214.8 &1145.5 &        & 3.276 \\
                       &        &        &       &        &       \\
\hline                                   
\end{tabular}
\end{table}

\newpage 

\centerline{\bf Figure captions}

\begin{itemize}

\item[{\bf Fig. 1}] $\Delta E_{1,\gamma}(I)$ (in keV), calculated from eq.
(1),
for octupole bands of a) $^{220}$Ra 
\cite{Ra220} and b) $^{224}$Ra \cite{Cocks1,Cocks2}. 
The experimental error in all cases is smaller than the symbol used 
for the points and therefore is not shown. (See table 1 for a list of the 
errors in the case of $^{220}$Ra.) 

\item[{\bf Fig. 2}] Odd--even staggering quantities 
$\Delta E_{1,\gamma}(I)$ (from eq. (1), in keV), 
$\Delta E_{45,E}(I)$ (from eq. (9), in keV), and $\Delta E_{NO}(I)$
(from eq. (14), in keV) for octupole bands of a) $^{220}$Ra \cite{Ra220}, 
b) $^{224}$Ra \cite{Cocks1,Cocks2}. 

\item[{\bf Fig. 3}] Same as fig. 1, but for a) $^{226}$Th \cite{Th226},
$^{228}$Th \cite{Th228}, $^{230}$Th \cite{Cocks2}, $^{232}$Th \cite{Cocks2}, 
$^{234}$Th \cite{Cocks2}, 
b) $^{230}$U \cite{U230}, $^{232}$U \cite{U232}, $^{234}$U \cite{U234}, 
$^{236}$U \cite{U236}, $^{238}$U \cite{U238}. 

\item[{\bf Fig. 4}] Same as fig. 1, but 
for octupole bands of a) $^{144}$Ba \cite{Hamilton}, b) $^{146}$Ba 
\cite{Hamilton}, c) $^{150}$Sm \cite{A150}, and d) $^{154}$Dy \cite{A154}. 

\item[{\bf Fig. 5}] Same as fig. 1, but for a) $^{152}$Sm \cite{A152},
$^{154}$Sm \cite{A154}, 
b) $^{152}$Gd \cite{A152}, $^{154}$Gd \cite{A154}, $^{156}$Gd \cite{A156}, 
$^{160}$Gd \cite{A160}, 
c) $^{156}$Dy \cite{Dy156}, $^{162}$Dy \cite{A162}, d) 
$^{162}$Er \cite{A162}, $^{164}$Er \cite{A164}. 

\end{itemize}

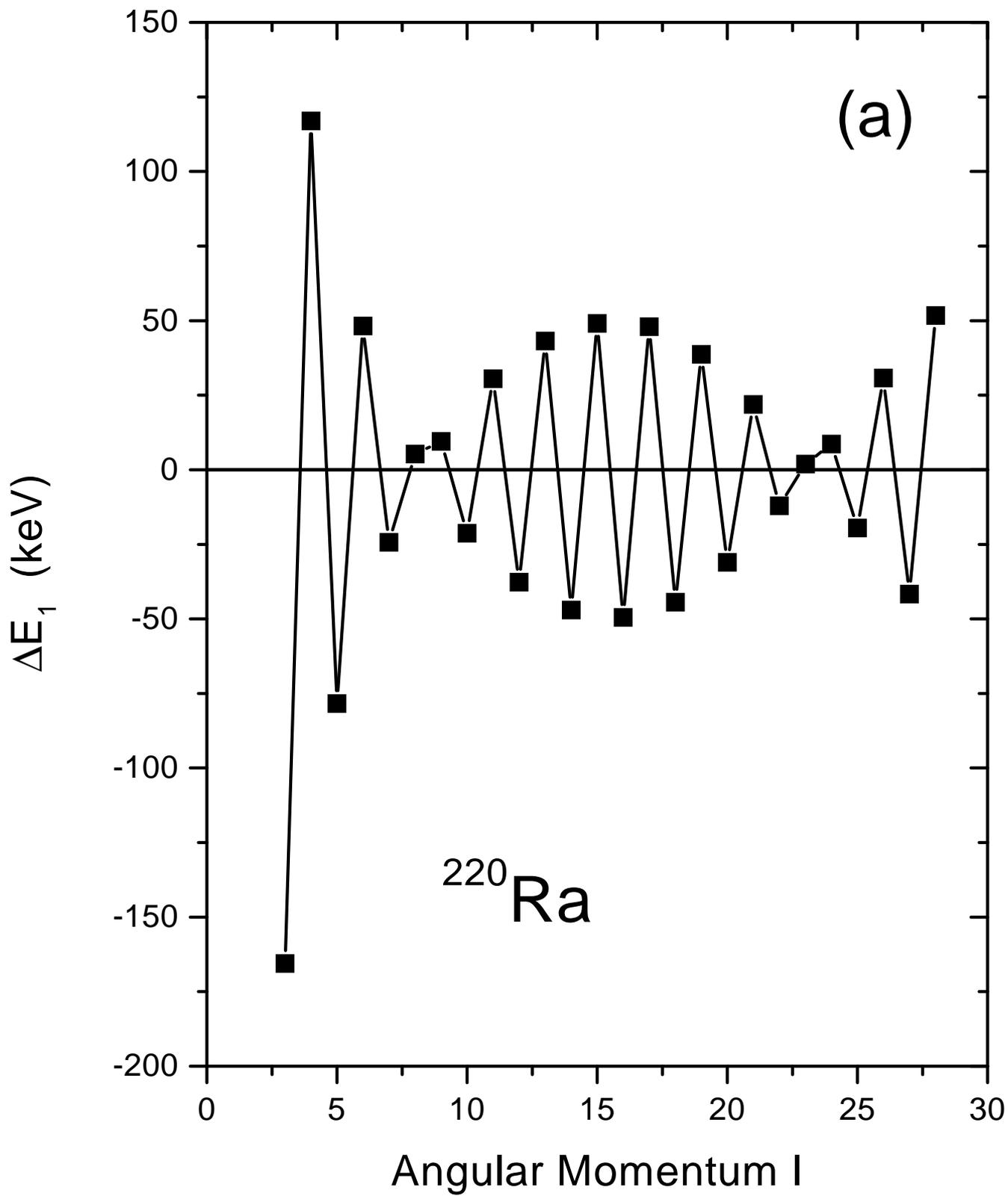
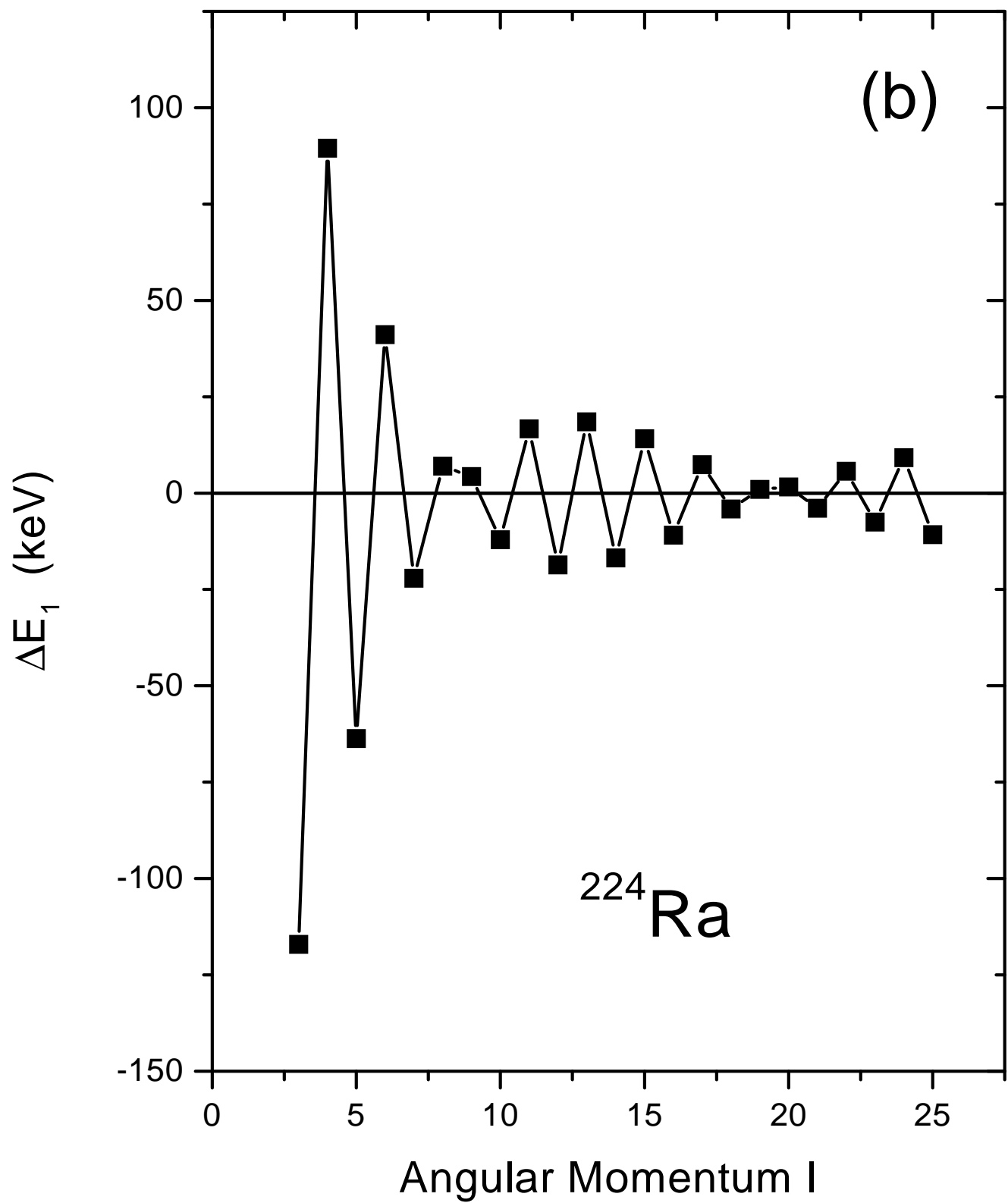
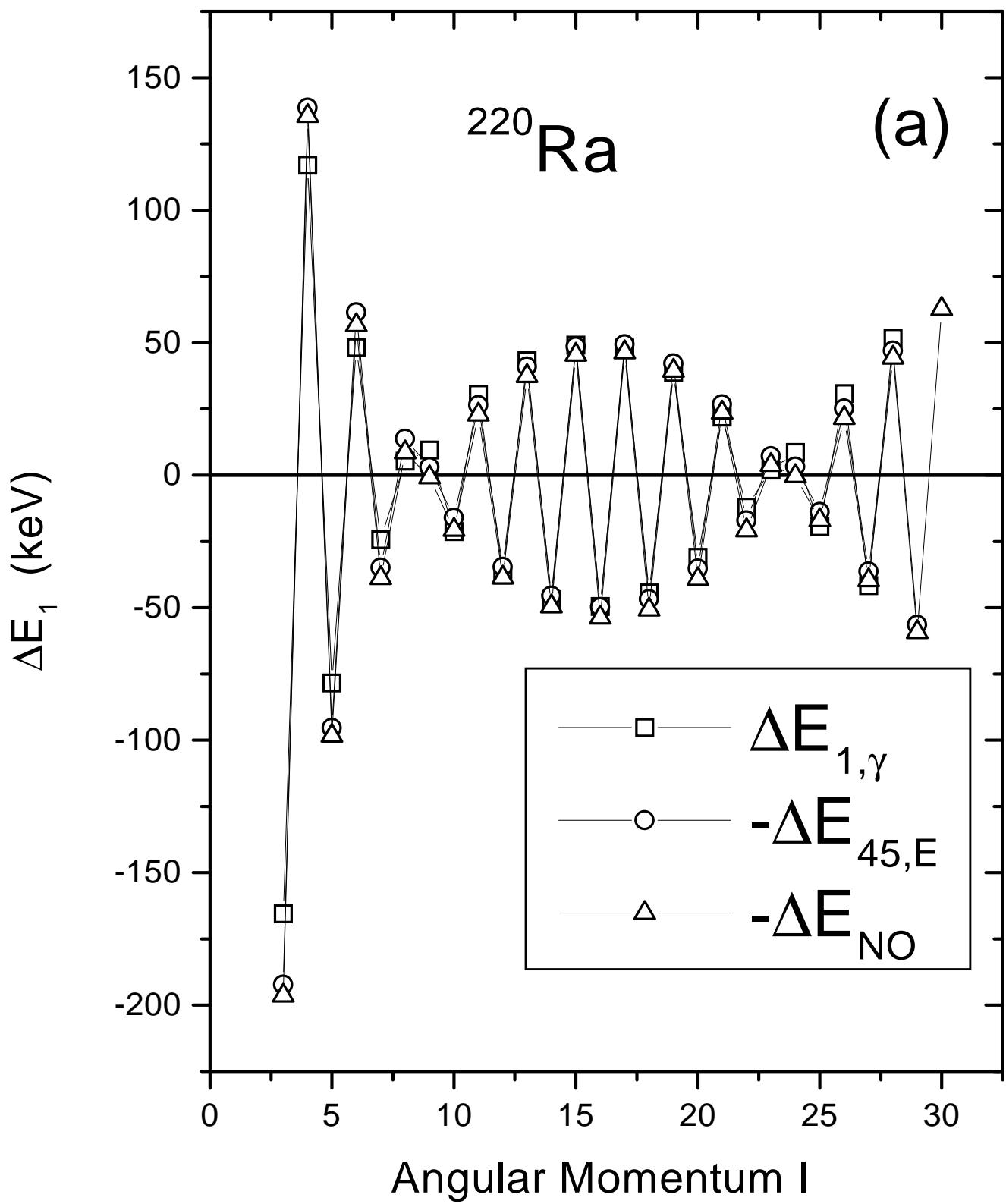
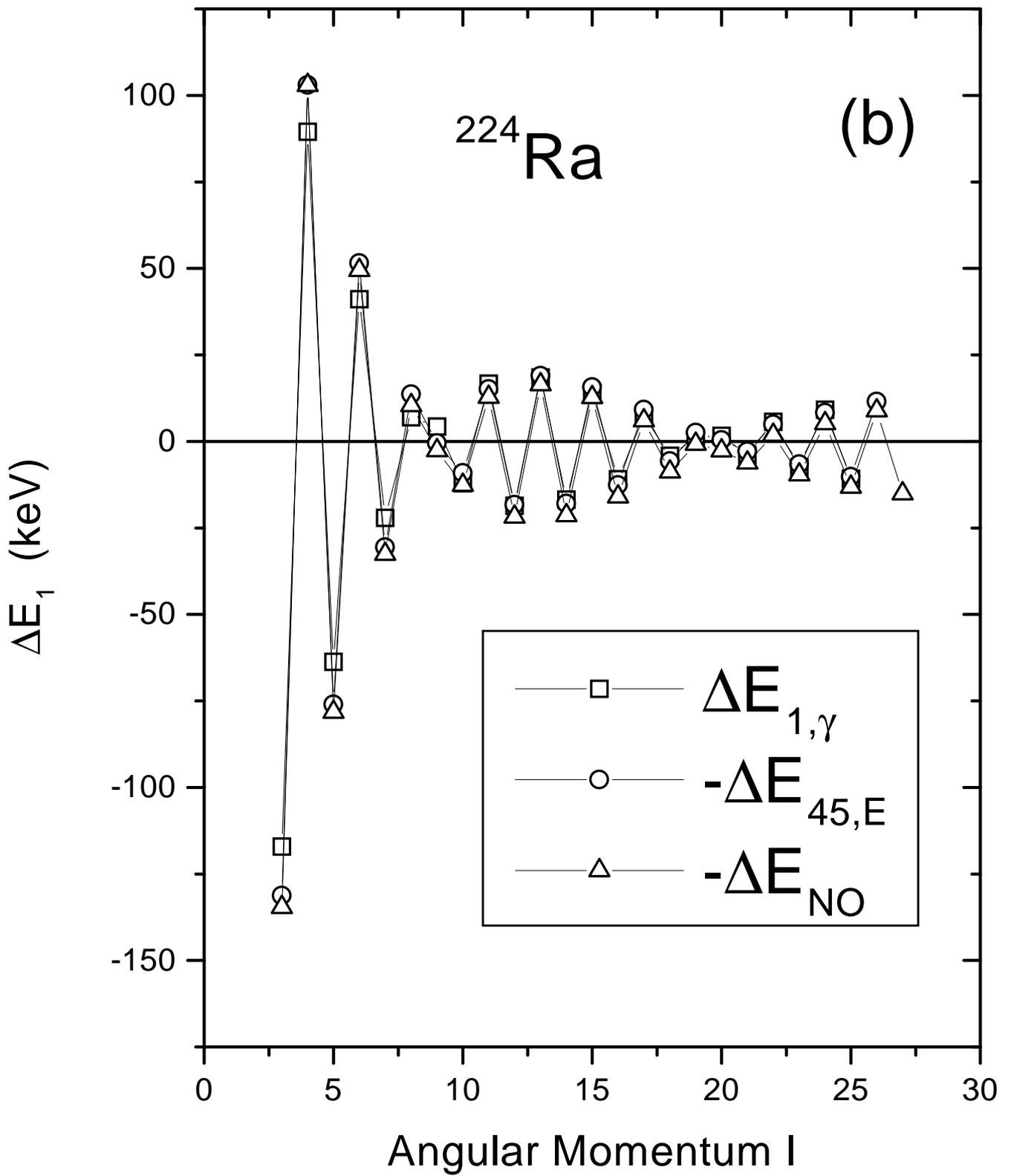
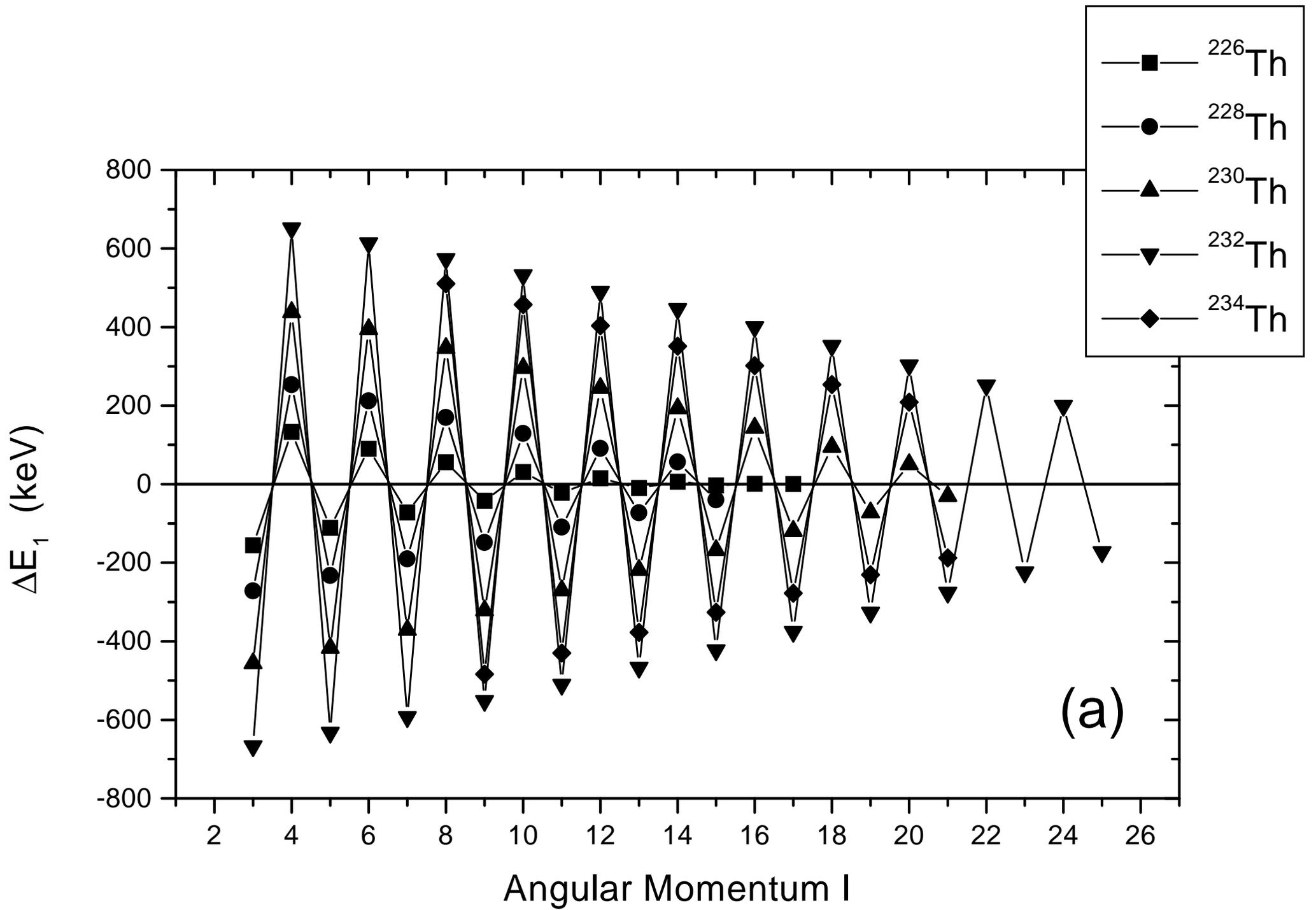
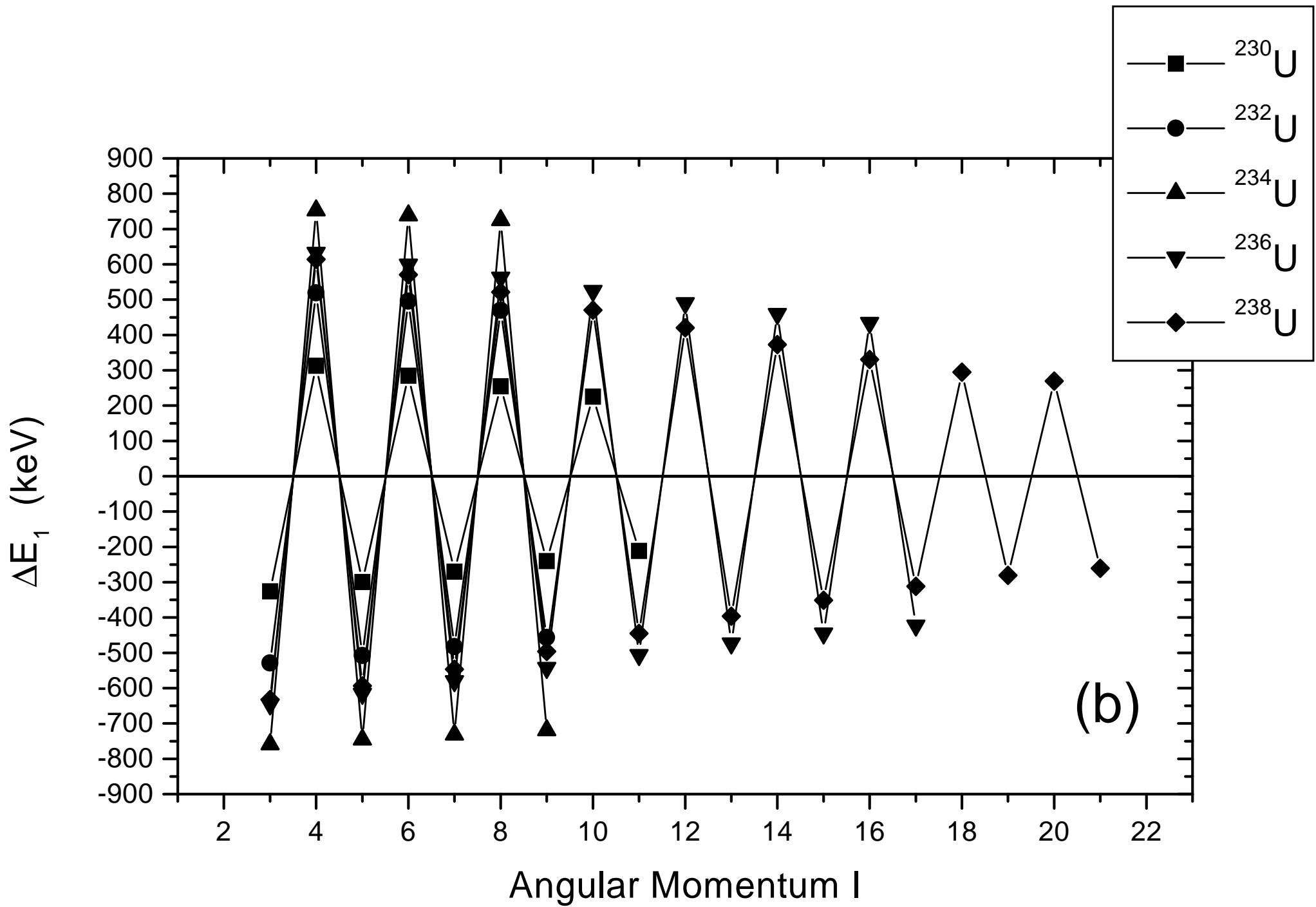
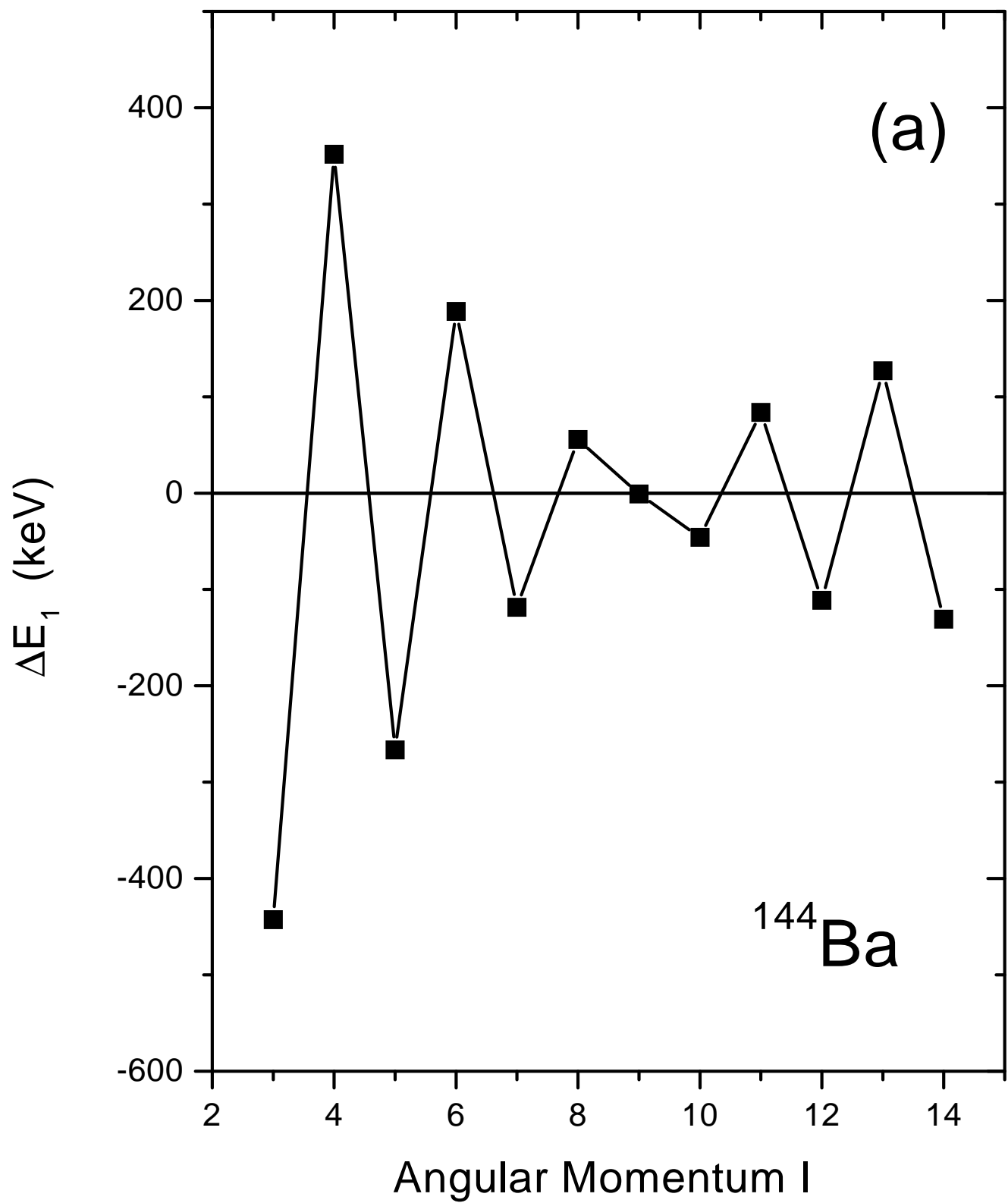
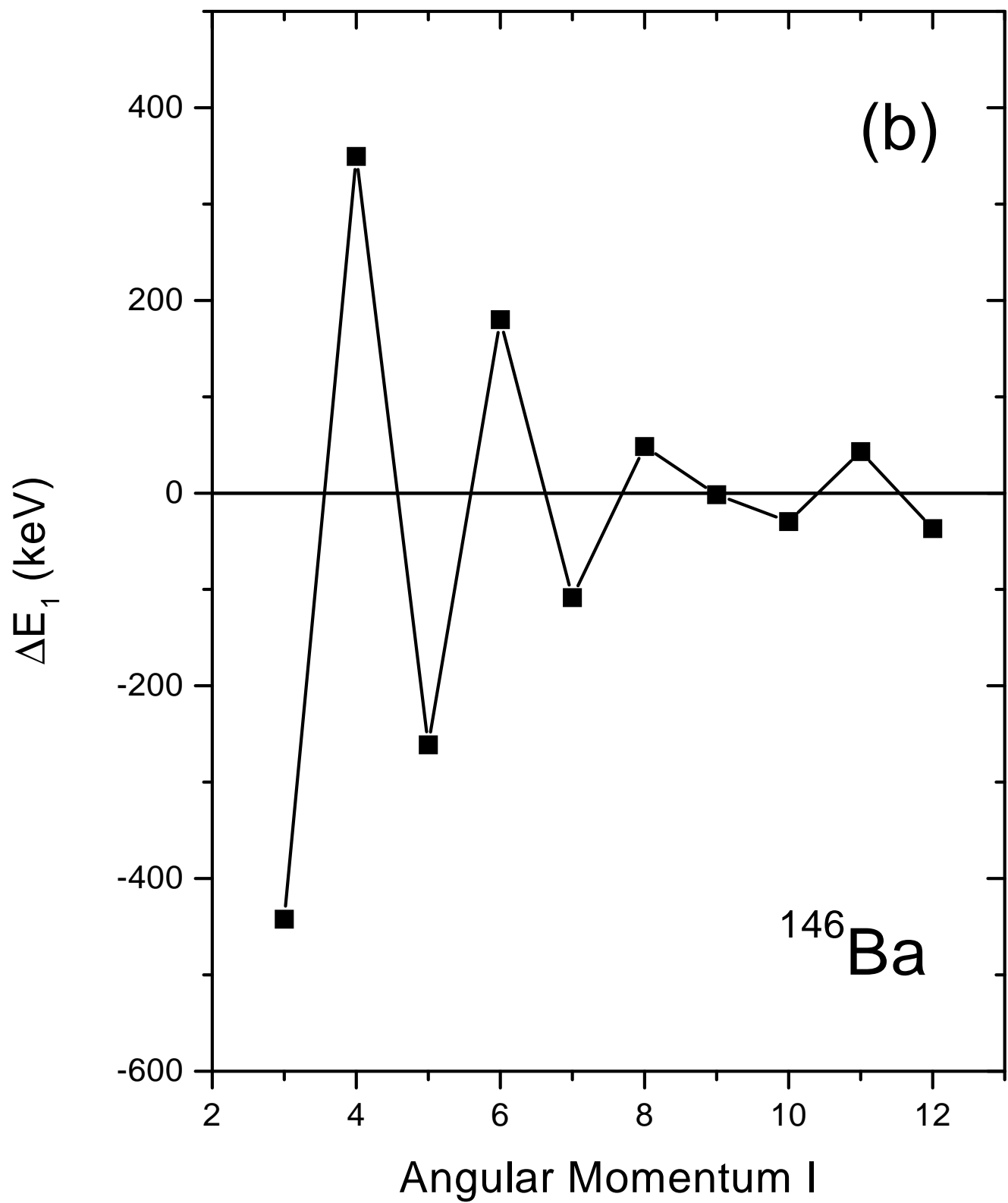
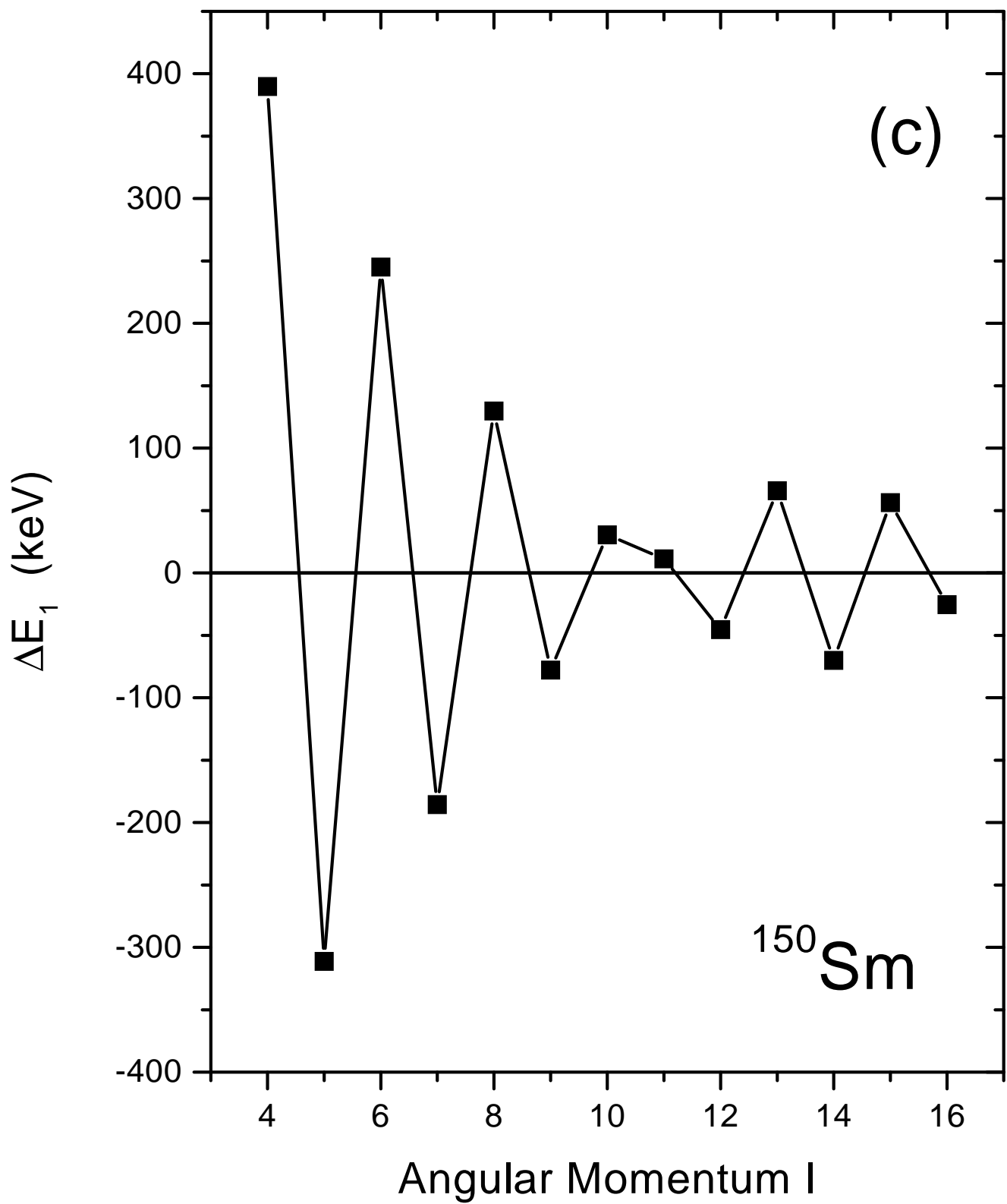
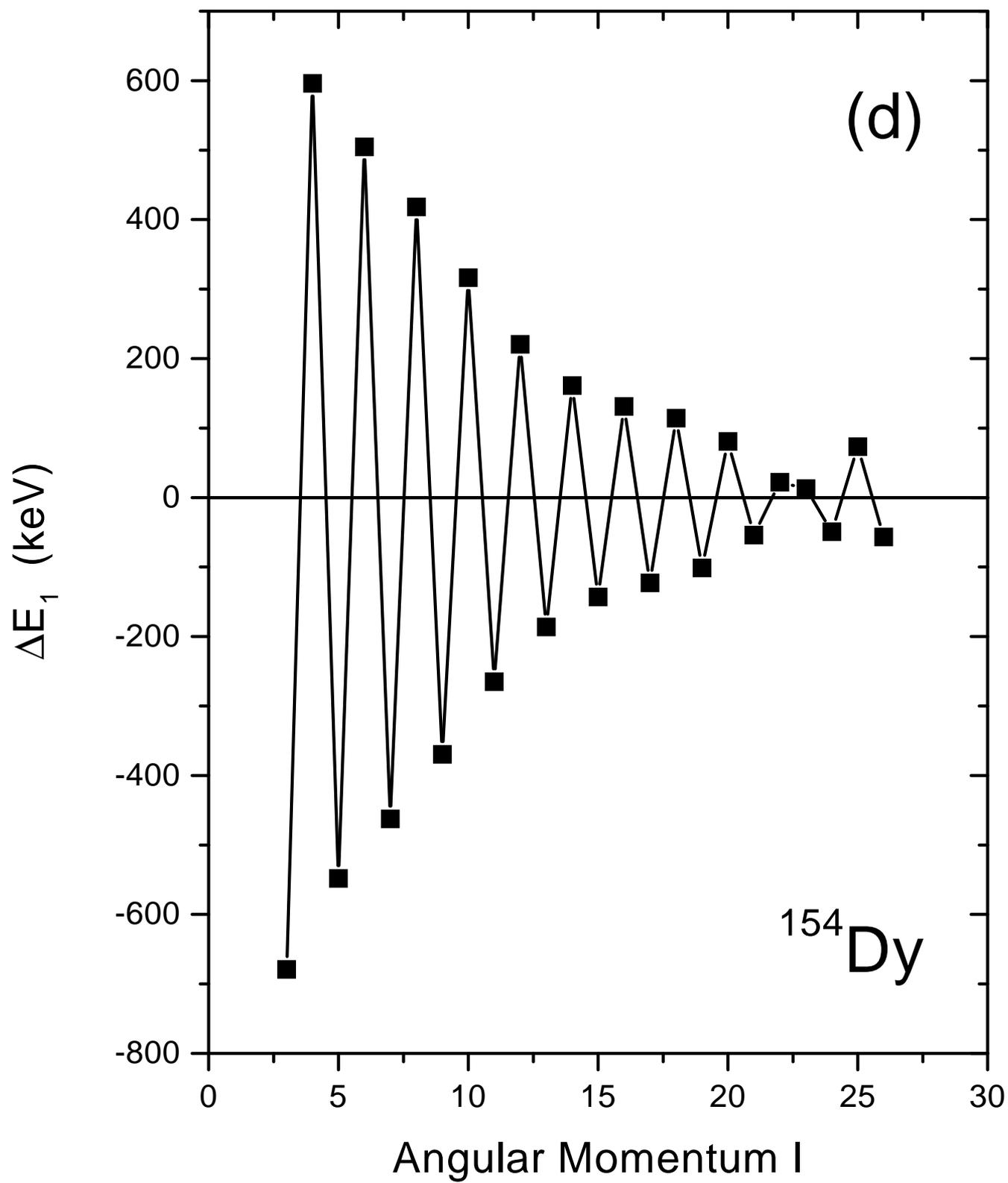
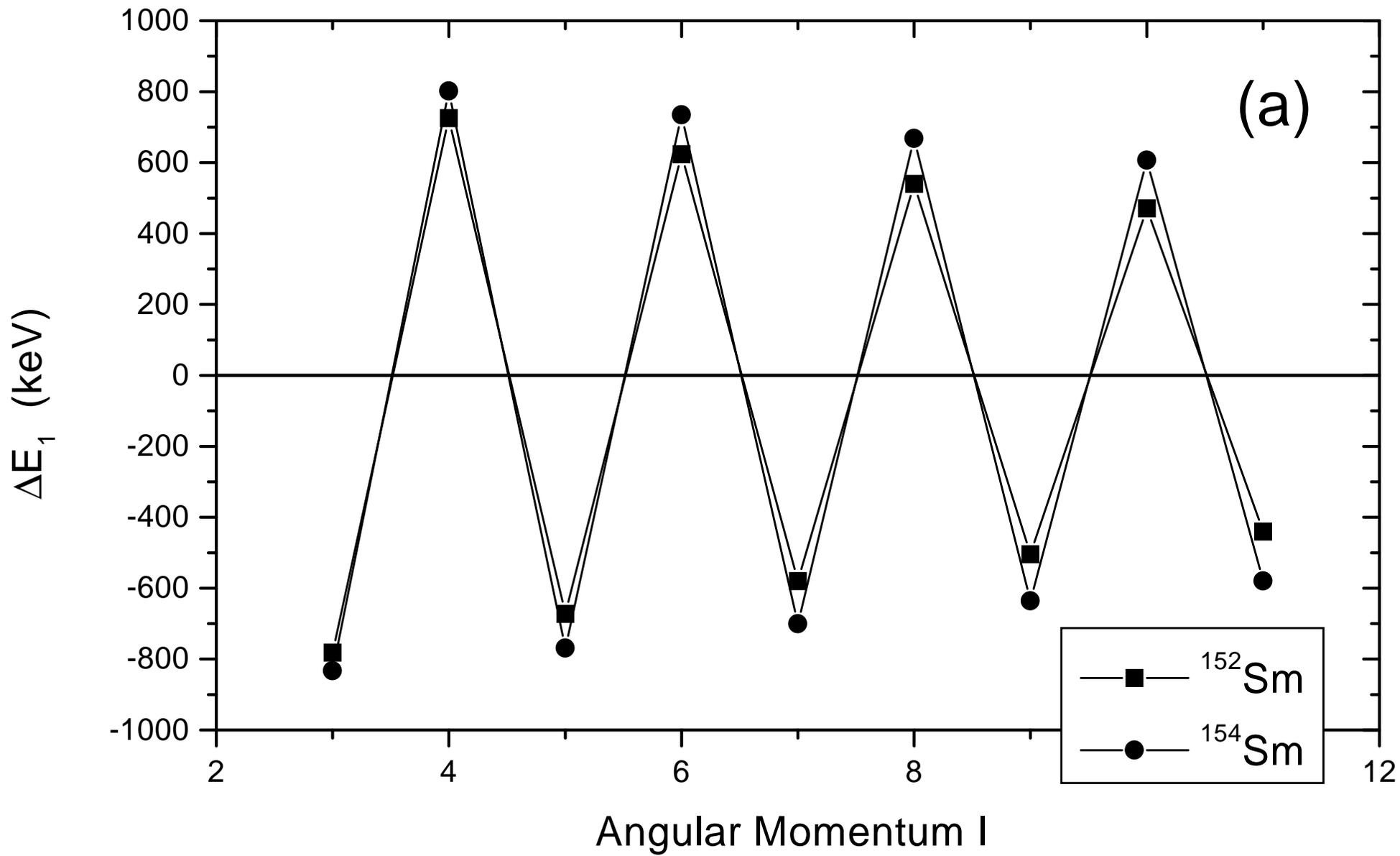
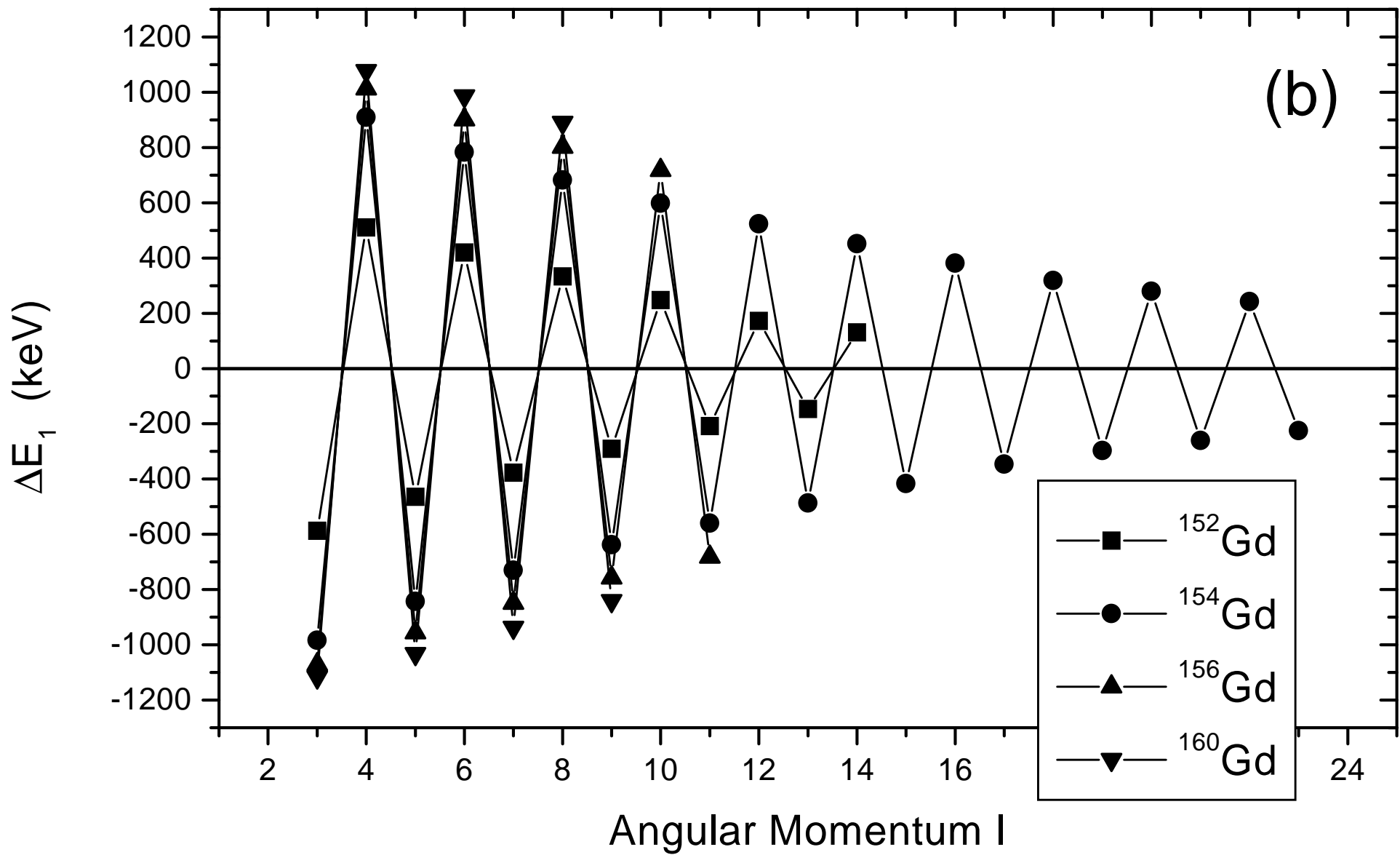
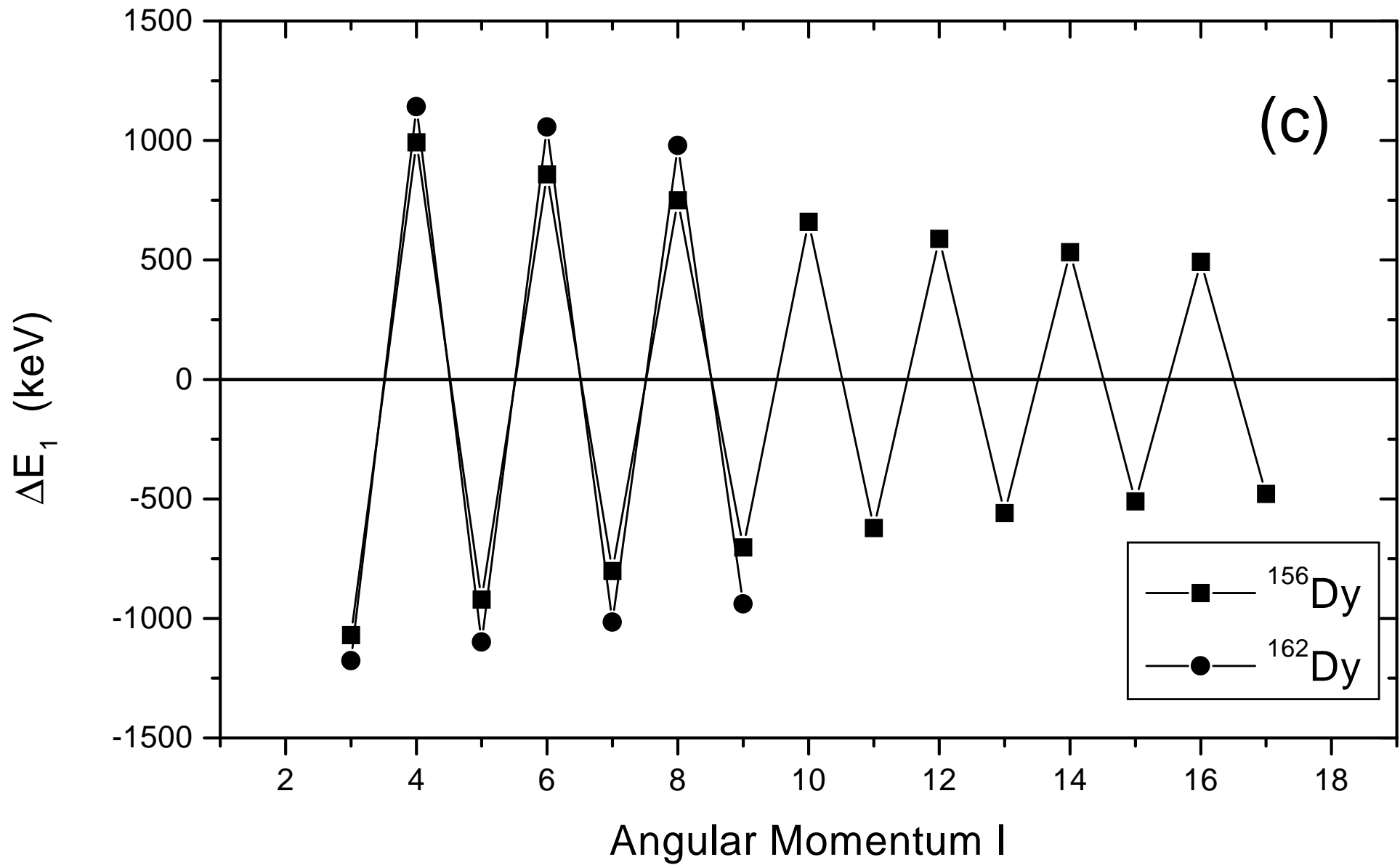
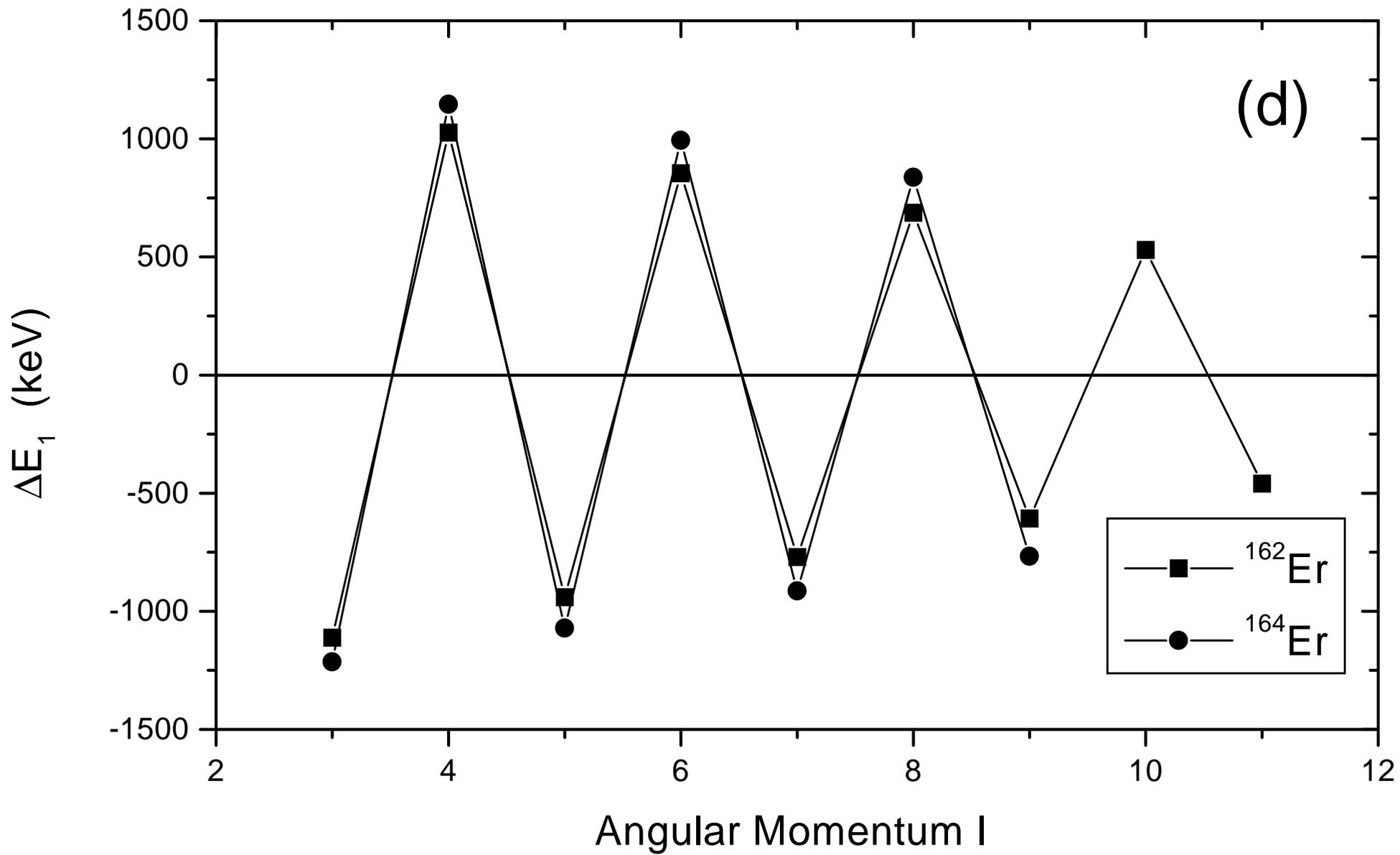
\end{document}